\documentclass{ws-ijns}
\usepackage{graphicx}
\usepackage[english]{babel}
\usepackage[super,sort,compress]{cite}
\usepackage{amsmath}
\usepackage{mathtools}

\usepackage{multirow}
\usepackage[flushleft]{threeparttable}
\usepackage[normalem]{ulem}

\usepackage{enumitem}
\newlist{tabitem}{itemize}{1}
\setlist[tabitem]{wide=0pt, nosep, leftmargin= * ,label=\textbullet,after=\vspace{-\baselineskip},before=\vspace{-2\baselineskip}}

\usepackage{hhline}

\usepackage[T1]{fontenc}
\usepackage{hyperref}

\begin{document}


\markboth{Peh et al.}{Automated Slowing EEG Classification}

\title{Multi-center validation study of automated classification of pathological slowing in adult scalp electroencephalograms via frequency features}

\author{Wei Yan Peh$^{1}$, 
John Thomas$^{1}$, 
Elham Bagheri$^{1}$,
Rima Chaudhari$^{2}$, 
Sagar Karia$^{3}$, 
Rahul Rathakrishnan$^{4}$, \\
Vinay Saini$^{5}$, 
Nilesh Shah$^{3}$,
Rohit Srivastava$^{5}$, 
Yee-Leng Tan$^{6}$,
and Justin Dauwels$^{1^{*}}$\footnote{Corresponding author, Email: jdauwels@ntu.edu.sg}}

\address{\vspace{0.5 cm} 
$^{1}$Nanyang Technological University, Singapore\\
$^{2}$Fortis Hospital Mulund, Mumbai, India\\
$^{3}$Lokmanya Tilak Municipal General Hospital, India\\
$^{4}$National University Hospital, Singapore\\
$^{5}$Department of Biosciences and Bioengineering, IIT Bombay, India\\
$^{6}$National Neuroscience Institute, Singapore}

\maketitle

\begin{abstract}
Pathological slowing in the electroencephalogram (EEG) is widely investigated for the diagnosis of neurological disorders. Currently, the gold standard for slowing detection is the visual inspection of the EEG by experts, which is time-consuming and subjective. To address those issues, we propose three automated approaches to detect slowing in EEG: Threshold-based Detection System (TDS), Shallow Learning-based Detection System (SLDS), and Deep Learning-based Detection System (DLDS). These systems are evaluated on channel-, segment-, and EEG-level. The three systems performs prediction via detecting slowing at individual channels, and those detections are arranged in histograms for detection of slowing at the segment- and EEG-level. We evaluate the systems through Leave-One-Subject-Out (LOSO) cross-validation (CV) and Leave-One-Institution-Out (LOIO) CV on four datasets from the US, Singapore, and India. The DLDS achieved the best overall results: LOIO CV mean balanced accuracy (BAC) of 71.9\%, 75.5\%, and 82.0\% at channel-, segment- and EEG-level, and LOSO CV mean BAC of 73.6\%, 77.2\%, and 81.8\% at channel-, segment-, and EEG-level. The channel- and segment-level performance is comparable to the intra-rater agreement (IRA) of an expert of 72.4\% and 82\%. The DLDS can process a 30-minutes EEG in 4 seconds and can be deployed to assist clinicians in interpreting EEGs.
\end{abstract}


\begin{multicols}{2}

\section{Introduction}
{S}{lowing} in electroencephalogram (EEG) is an indication of potential neurological dysfunctions such as epilepsy, stroke, or dementia \cite{sheorajpanday2011quantitative, kaszniak1979cerebral, donnelly2007focal, britton2016electroencephalography}. An abnormal amount of slowing in EEG suggests neurological abnormalities or poor prognosis for neurological recovery. The severity of slowing is dependent on EEG frequency (delta or theta), duration (continuous or intermittent), and location of the slowing (focal or generalized) \cite{britton2016electroencephalography}. Slow wave activity usually appears under 4Hz in the delta band region but can also appear in the theta band region (between 4 to 8Hz) \cite{soikkeli1991slowing, holler2015there}.

When slowing occurs in 90\% or more in the EEG recording, it is considered continuous slowing. Otherwise, it is considered intermittent slowing if it occurs between 20\% and 90\% of the recording \cite{kane2017revised, tatum2014handbook}. Meanwhile, generalized slowing occurs throughout the brain, whereas focal slowing occurs only in one part(s) of the brain \cite{donnelly2007focal}. Generalized and continuous slowing often leads to a poorer prognosis for recovery \cite{godefroy2010stroke, nadlonek2015early}. However, it is normal to observe slow waves in EEG during drowsiness, sleep, or hyperventilation. They also can appear in EEG for the elderly as a slow background or Posterior Slow Waves of Youth (PSWY) in adolescence \cite{choi2019resting, ohoyama2012source}. Additionally, EEG artifacts such as excessive sweating, muscle movement, or eye movement can look similar to EEG slowing \cite{islam2016methods, kalevo2019improved}. Hence, the classification of pathological slowing can be challenging.

In current clinical practice, the gold standard for slowing annotation in EEG is through visual inspection by neurologists. This can be a time-consuming process. Moreover, slowing annotation can be strenuous due to the variation in the slowing duration and location. Hence, there is a need for an automated slowing detection system to hasten the process.

As far as we know, no studies have investigated how to detect slowing from EEG in an automated manner directly. Instead, existing methods aim to detect neurological disorders from EEG that exhibit slowing (such as stroke, brain injury, seizures), without detecting EEG slowing explicitly \cite{aminov2017acute, iyer2017effective, bentes2018quantitative, chen2018transcranial}. Spectral features are widely applied for such analysis as they are scale-invariant (independent of amplitude or power) \cite{riaz2015emd}. For EEG classification, the methods adopted are simple thresholding, traditional machine (shallow) learning, or deep learning via Convolutional Neural Networks (CNN) \cite{sheorajpanday2011quantitative, wang2014emotional, finnigan2016defining, thomas2017deep, thomas2018eeg, ansari2019neonatal, roy2019chrononet, thomas2021automated}. One drawback of most CNN approaches is that they only investigate single-channel EEG or only assess the CNNs on a small set of multi-channel EEGs from a single institution \cite{sors2018convolutional, michielli2019cascaded, mousavi2019deep, supratak2017deepsleepnet}. Additionally, current approaches do not explicitly detect pathological slowing. Instead, they classify EEGs with neurological conditions directly \cite{gemein2020machine}.

An abnormal quantity of slowing in EEG provides information to diagnose a neurological disorder and prescribe an appropriate and timely treatment. For instance, brain tumors may be associated with localized and continuous EEG slowing \cite{handayani2019eeg}. Hence, identifying brain regions that exhibit that type of slowing may help confirm or better localize brain tumors. Consequently, there is a demand for automated EEG classification systems that detect abnormal slowing in EEGs for a more reliable diagnosis. To address these shortcomings, we design three automated systems to detect pathological slowing and evaluated the systems on EEG datasets from multiple institutions in Singapore, India, and the US. 

\begin{figurehere}
\centering
\includegraphics[width=8cm]{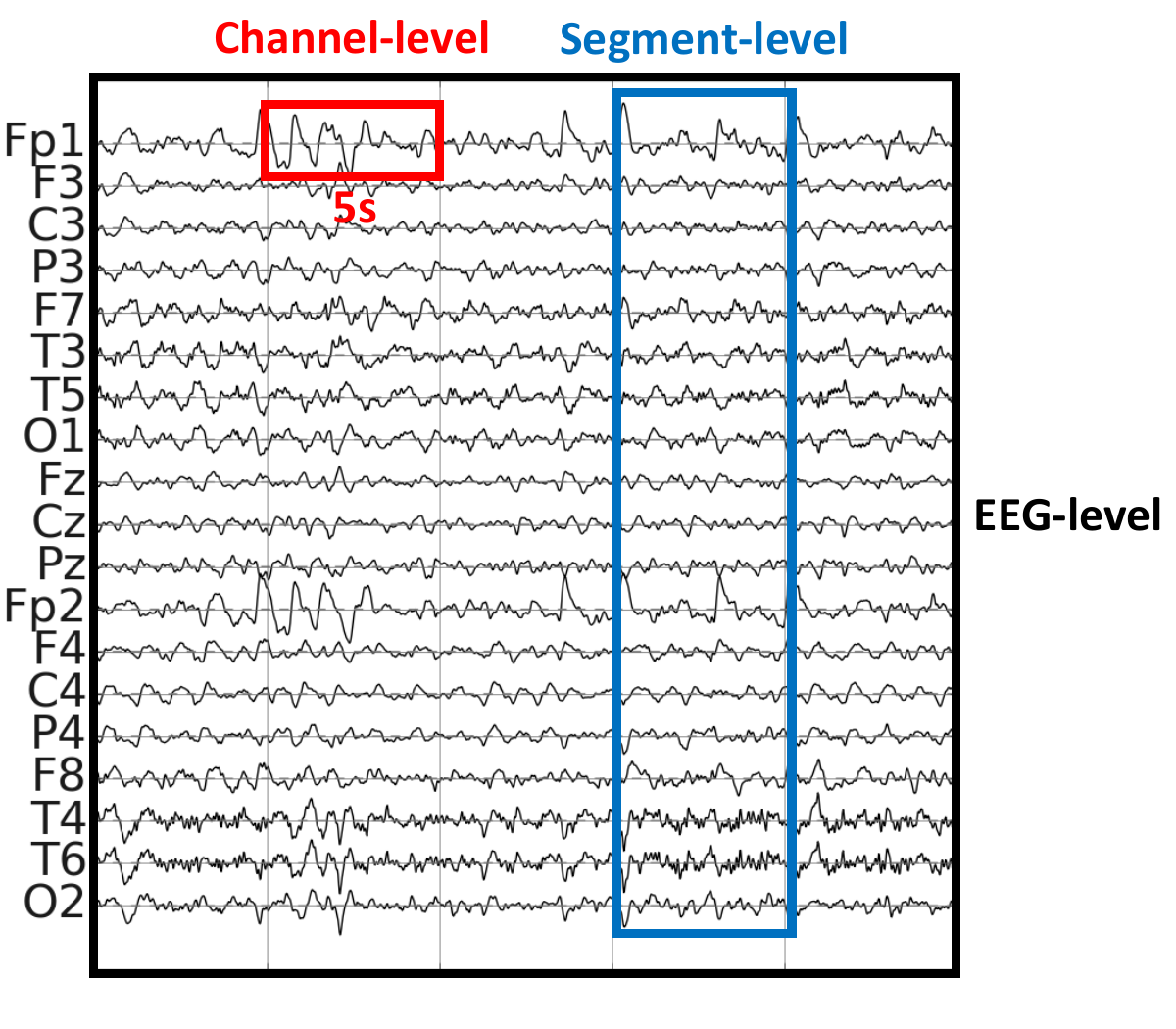}
\vspace{-1.0 cm}
\caption{Channel-, segment- and EEG-level.}
\label{fig:classification_level}
\vspace{-0.4 cm}
\end{figurehere}

In this paper, we proposed three automated systems for detecting pathological slowing in EEG (see Table \ref{tab:EEG_system_summary}): Threshold-based Detection System (TDS), Shallow Learning-based Detection System (SLDS), and Deep Learning-based Detection System (DLDS). The systems detect slowing at single-channel EEG segments (channel-level), multi-channel segments (segment-level), and full EEGs (EEG-level), allowing us to detect slowing at all scales (see Figure \ref{fig:classification_level}). The TDS performs segment- and EEG-level classification without a trained machine learning or deep learning-based channel-level EEG slowing detector. Instead, it uses simple thresholding. By contrast, the SLDS and DLDS perform the classification in two stages. The first stage deploys a shallow learning or deep learning-based slowing detector to detect slowing at the channel-level. The second stage utilizes the channel-level detections to identify slowing in EEG segments or full EEGs.

The TDS was designed as a benchmark to assess the improvement afforded by a system with a machine learning or deep learning-based channel-level slowing detector. Meanwhile, we implemented a shallow and deep learning model for the channel-level slowing detector to quantify the advantages of deep learning. To the best of our knowledge, this current study is the first to design channel-, segment-, and EEG-level slowing classification systems. Additionally, this study is the first to detect pathological slowing in EEG without information about the potentially underlying neurological disorder. As far as we know, all prior studies concentrate on detecting neurological disorders from EEG, which usually also generate EEG slowing. The channel-level slowing detector is implemented to detect slowing at single-channel EEG channels. Meanwhile the segment- and EEG-level slowing classifier are implemented to illustrate that our systems are capable of detecting abnormal amount of slowing in EEGs of various lengths, from a short 5s multi-channel EEG segment to multi-channel routine EEG recordings of 30min or longer.

We validate the performance of the proposed systems on multiple datasets by considering two real-world scenarios. In the first scenario, we assume to have access to some past EEGs (around 50 to 100 EEGs) and their clinical reports. With the data, we can retrain the classification system to perform predictions on EEGs from other patients from the same center in the future. To assess the performance of the system in this scenario, we apply Leave-One-Subject-Out (LOSO) cross-validation (CV) for each institute (dataset) separately. In LOSO CV, we select one subject for testing and the remaining subjects to train the classification system. We repeat this for each subject and compute the performance of the systems across all the subjects.

In the second scenario, we assume that no EEGs nor clinical reports are available from the new center for calibration. Instead, we utilize existing datasets to train the classification system to predict the labels of those EEGs from the new center. We evaluate our proposed systems under this scenario by Leave-One-Institution-Out (LOIO) CV. First, we select an institute of our pool of participating institutes (cf. Section 2.1) and leave it out for testing. The EEGs from the remaining institutes are employed to train the classification system. We repeat this for each institution. 

To the best of our knowledge, this current study is the first to perform a cross-institutional assessment of automated EEG classification systems to detect pathological slowing. It is crucial to perform the LOIO CV assessment to evaluate the generalizability of the proposed system. Similarly, we perform the LOSO CV assessment to evaluate the proposed classification systems after recalibration for a particular dataset.

As our results show, the DLDS achieves the best overall performance for the three classification tasks. It yields an LOIO CV mean balanced accuracy (BAC) of 71.9\%, 75.5\%, and 82.0\%, for the channel-, segment- and EEG-level classification, respectively, whereas the LOSO CV mean BAC are 73.6\%, 77.2\%, and 81.8\%, respectively. The channel- and segment-level intra-rater agreement (IRA) of an expert is 72.4\% and 82\% respectively on the same data. Thus, the DLDS can detect abnormal slowing in channels and segments reliably at the human expert level. Moreover, the DLDS can process a 30-minute EEG in about 4 seconds. Therefore, the proposed systems for automated detection of EEG slowing might be useful in clinical applications.

The rest of this paper is organized as follows. In Section~\ref{dataset}, we describe the EEG datasets and the preprocessing steps employed. In Section~\ref{feature}, we review various spectral features considered, while in Section~\ref{annotation}, we describe the EEG channel and segment datasets. In Section~\ref{classification_systems}, we present the three proposed machine learning systems for channel-, segment-, and EEG-level slowing detection. In Section~\ref{results}, we show numerical results for the classification systems and tasks, while in Section~\ref{discuss}, we discuss the performance of the proposed systems and their potential relevance in clinical practice. Lastly, in Section~\ref{conclude}, we offer concluding remarks and suggestions for future work.

\begin{table*}[htb]
\centering
\tbl{Summary of the three EEG classification system. \label{tab:EEG_system_summary}}        
{
\scalebox{0.90}{
\begin{threeparttable}
\begin{tabular}{|c|p{6cm}|p{8cm}|} 
\hline
\begin{tabular}[c]{@{}c@{}} \textbf{}\textbf{Classification}\\\textbf{System} \end{tabular} 
& 
\multicolumn{1}{c|}{\begin{tabular}[c]{@{}c@{}}\textbf{Channel-Level}\\\textbf{Slow Detection} \end{tabular}} 
& 
\multicolumn{1}{c|}{\begin{tabular}[c]{@{}c@{}}\textbf{Segment- or EEG-Level}\\\textbf{Slow Detection}\end{tabular}} 
\\ 
\hline

\begin{tabular}[c]{@{}c@{}}
\textbf{Threshold-based}\\\textbf{Detection}\\\textbf{System}\\\textbf{(TDS)}\end{tabular}
&  
\begin{tabitem}
    \item \textbf{Features}: Spectral features
    \item \textbf{Classifier}: Simple thresholding
\end{tabitem}
&  
\begin{tabitem}
    \item \textbf{Features}: Histogram-based features$^{*}$ of the spectral measures computed from 5s single-channel segments (with 75\% overlap)
    \item \textbf{Segment/EEG Classifier}: Shallow learning model
\end{tabitem}
\\ 
\hline

\begin{tabular}[c]{@{}c@{}} 
\textbf{Shallow Learning-based}\\ 
\textbf{Detection}\\
\textbf{System}\\ 
\textbf{(SLDS)}\end{tabular} 
&  
\begin{tabitem}
    \item \textbf{Features}: Spectral features
    \item \textbf{Classifier}: Shallow learning model
\end{tabitem}
& 
\begin{tabitem}
    \item \textbf{Features}: Histogram-based features$^{*}$ of channel-level detector outputs computed from 5s single-channel segments (with 75\% overlap)
    \item \textbf{Segment/EEG Classifier}: Shallow learning model
\end{tabitem}
\\ 
\hline

\begin{tabular}[c]{@{}c@{}} 
\textbf{Deep Learning-based}\\ 
\textbf{Detection}\\ 
\textbf{System}\\ 
\textbf{(DLDS)}\end{tabular} 
&  
\begin{tabitem}
    \item \textbf{Features}: EEG spectrum
    \item \textbf{Classifier}: Deep learning model (CNN)
\end{tabitem}
& 
\begin{tabitem}
    \item \textbf{Features}: Histogram-based features$^{*}$ of channel-level detector outputs computed from 5s single-channel segments (with 75\% overlap)
    \item \textbf{Segment/EEG Classifier}: Shallow learning model
\end{tabitem}
\\
\hline

\end{tabular}

\begin{tablenotes}
\setlength\labelsep{0pt}
\footnotesize
\item $^{*}$ Histogram counts, and the mean, median, mode, standard deviation (std), minimum value, maximum value, range, kurtosis, and skewness.
\end{tablenotes}

\end{threeparttable}
}}
\vspace{-0.2 cm}
\end{table*}

\section{Materials and Methods}
\subsection{Scalp EEG dataset}
\label{dataset}
We analyzed scalp EEG recording from five institutions:

\vspace{-0.3 cm}
\begin{enumerate}[leftmargin=*,align=left]
  \item Temple University Hospital (TUH), USA.
  \item National Neuroscience Institute (NNI), Singapore.
  \item National University Hospital (NUH), Singapore.
  \item Fortis Hospital, Mumbai, India.
  \item Lokmanya Tilak Municipal General Hospital (LTMGH), Mumbai, India.
\end{enumerate}
\vspace{-0.1 cm}

The review boards of the respective institutions have approved this study. The EEGs were recorded by 19 electrodes placed according to the International 10-20 System. The datasets predominantly consist of awake adult EEGs (age$\geq$18 years). We have access to the EEGs and their clinical reports, except for the NUH dataset. Hence, we cannot perform EEG-level classification as we have no access to information about slowing in the NUH dataset.

If an EEG report mentions abnormal slowing, we assume that the corresponding EEG indeed contains pathological slowing; otherwise, the EEG is considered free of slowing. The proposed EEG-level classifiers aim to predict whether pathological slowing is mentioned in the clinical report for an EEG. The details of the EEG datasets are tabulated in Table \ref{tab:patient_age_gender}.

\begin{table*}[htb]
\centering
\tbl{Patient information for the different EEG datasets. \label{tab:patient_age_gender}}        
{
\scalebox{0.85}{
\begin{threeparttable}

\begin{tabular}{|c|c|ccccc|ccccc|} 
\cline{3-12}
\multicolumn{1}{c}{} &  & \multicolumn{5}{c|}{\textbf{Slow-free EEG}} & \multicolumn{5}{c|}{\textbf{Slowing EEG}} \\ 
\hline
\textbf{Dataset ($F_{s}$)}  & \begin{tabular}[c]{@{}c@{}}\textbf{Total}\\\textbf{EEG} \end{tabular} & \textbf{EEG}  & \textbf{Gender}  & \textbf{No}  & \begin{tabular}[c]{@{}c@{}}\textbf{Duration}\\\textbf{(minutes)} \end{tabular} & \begin{tabular}[c]{@{}c@{}}\textbf{Age}\\\textbf{(years)} \end{tabular} & \textbf{EEG}  & \textbf{Gender}  & \textbf{No}  & \begin{tabular}[c]{@{}c@{}}\textbf{Duration}\\\textbf{(minutes)} \end{tabular} & \begin{tabular}[c]{@{}c@{}}\textbf{Age}\\\textbf{(years)} \end{tabular} \\ 
\hline
\multirow{2}{*}{\begin{tabular}[c]{@{}c@{}}\textbf{TUH}\\\textbf{(250, 256, 500 Hz)} \end{tabular}} & \multirow{2}{*}{141} & \multirow{2}{*}{99} & M & 46 & 22.19$\pm$4.36  & 42.02$\pm$14.44  & \multirow{2}{*}{42} & M & 28 & 11.58$\pm$6.07  & 52.96$\pm$10.39  \\
 &  &  & F & 53 & 21.26$\pm$2.03  & 46.17$\pm$16.87  &  & F & 14 & 19.74$\pm$4.09  & 47.5$\pm$18.62  \\ 
\hline
\multirow{2}{*}{\begin{tabular}[c]{@{}c@{}}\textbf{NNI}\\\textbf{(200 Hz)} \end{tabular}} & \multirow{2}{*}{114} & \multirow{2}{*}{58} & M & 29 & 27.78$\pm$0.64  & 45.62$\pm$17.27  & \multirow{2}{*}{56} & M & 25 & 27.64$\pm$1.58  & 51.16$\pm$18.35  \\
 &  &  & F & 29 & 27.43$\pm$1.95  & 52.31$\pm$19.87  &  & F & 31 & 28.04$\pm$1.29  & 52.94$\pm$19.73  \\ 
\hline
\multirow{3}{*}{\begin{tabular}[c]{@{}c@{}}\textbf{Fortis}\\\textbf{(500 Hz)} \end{tabular}} & \multirow{3}{*}{358} & \multirow{3}{*}{285} & M & 155 & 20.87$\pm$6.53  & 45.86$\pm$19.69  & \multirow{3}{*}{73} & M & 50 & 20.3$\pm$2.95  & 55.52$\pm$17.83  \\
 &  &  & F & 123 & 20.26$\pm$4.07  & 45.74$\pm$18.23  &  & F & 19 & 20.61$\pm$3.54  & 50.0$\pm$16.92  \\
 &  &  & UNK & 7 & 20.68$\pm$1.03  & 43.0$\pm$17.86  &  & UNK & 4 & 22.16$\pm$1.52  & 63.75$\pm$5.26  \\ 
\hline
\multirow{2}{*}{\begin{tabular}[c]{@{}c@{}} \textbf{LTMGH}\\\textbf{(256 Hz)} \end{tabular}} & \multirow{2}{*}{1100} & \multirow{2}{*}{701} & M & 370 & 14.01$\pm$1.49  & 33.49$\pm$18.29  & \multirow{2}{*}{399} & M & 207 & 14.77$\pm$1.88  & 37.03$\pm$24.26  \\
 &  &  & F & 331 & 14.27$\pm$1.73  & 31.04$\pm$18.7  &  & F & 192 & 14.65$\pm$2.61  & 36.8$\pm$21.79  \\ 
\hline \hline
\multirow{3}{*}{\textbf{All}} & \multirow{3}{*}{1713} & \multirow{3}{*}{1143} & M & 600 & 17.08$\pm$5.56  & 37.93$\pm$19.22  & \multirow{3}{*}{570} & M & 310 & 16.41$\pm$4.88  & 42.59$\pm$23.33  \\
 &  &  & F & 536 & 17.05$\pm$4.58  & 37.06$\pm$20.06  &  & F & 256 & 16.99$\pm$5.24  & 40.32$\pm$21.95  \\
 &  &  & UNK & 7 & 20.68$\pm$1.03  & 43.0$\pm$17.86  &  & UNK & 4 & 22.16$\pm$1.52  & 63.75$\pm$5.26  \\ 
\hline
\multicolumn{1}{c}{} & \multicolumn{1}{l}{} & \multicolumn{1}{l}{} & \multicolumn{1}{l}{} & \multicolumn{1}{l}{} & \multicolumn{1}{l}{} & \multicolumn{1}{l}{} & \multicolumn{1}{l}{} & \multicolumn{1}{l}{} & \multicolumn{1}{l}{} & \multicolumn{1}{l}{} & \multicolumn{1}{l}{} \\ 
\cline{3-7}
\multicolumn{1}{l}{} & \multicolumn{1}{l|}{} & \multicolumn{5}{c|}{\textbf{All EEG}} & \multicolumn{1}{l}{} & \multicolumn{1}{l}{} & \multicolumn{1}{l}{} & \multicolumn{1}{l}{} & \multicolumn{1}{l}{} \\ 
\cline{1-7}
\textbf{Dataset ($F_{s}$)}  & \begin{tabular}[c]{@{}c@{}}\textbf{Total}\\\textbf{EEG} \end{tabular} & \textbf{EEG}  & \textbf{Gender}  & \textbf{No}  & \begin{tabular}[c]{@{}c@{}}\textbf{Duration}\\\textbf{(minutes)} \end{tabular} & \begin{tabular}[c]{@{}c@{}}\textbf{Age}\\\textbf{(year)} \end{tabular} & \multicolumn{1}{l}{} & \multicolumn{1}{l}{} & \multicolumn{1}{l}{} & \multicolumn{1}{l}{} & \multicolumn{1}{l}{} \\ 
\cline{1-7}
\multirow{2}{*}{\begin{tabular}[c]{@{}c@{}}\textbf{NUH}\\\textbf{(250 Hz)} \end{tabular}} & \multirow{2}{*}{150} & \multirow{2}{*}{150} & M & 89 & 19.36$\pm$9.36  & 51.23$\pm$19.91  & \multicolumn{1}{l}{} & \multicolumn{1}{l}{} & \multicolumn{1}{l}{} & \multicolumn{1}{l}{} & \multicolumn{1}{l}{} \\
 &  &  & F & 61 & 19.60$\pm$9.30  & 56.48$\pm$20.18  & \multicolumn{1}{l}{} & \multicolumn{1}{l}{} & \multicolumn{1}{l}{} & \multicolumn{1}{l}{} & \multicolumn{1}{l}{} \\
\cline{1-7}
\end{tabular}

\begin{tablenotes}
\setlength\labelsep{0pt}
\footnotesize
\item $F_s$: sampling frequency, M: male, F: female, UNK: unknown, age/duration are reported as mean $\pm$ std.
\item Note: The NUH dataset does not have slowing labeled in the clinical report.
\end{tablenotes}

\end{threeparttable}
}}
\vspace{-0.25 cm}
\end{table*}

The TUH dataset is the largest public epilepsy EEG dataset. Concretely, we investigate two corpora from the TUH dataset: TUH Slowing Corpus and TUH Abnormal Corpus \cite{obeid2016temple}. The NNI, Fortis, and NUH datasets consist of scalp EEGs recorded during routine clinical care. However, the clinical reports are unavailable for the NUH dataset; hence, the NUH dataset is only deployed for the segment- and channel-level annotation.

Similarly, the LTMGH dataset consists of routine scalp EEGs. However, unlike the other datasets, the LTMGH EEGs were recorded by EEG recording equipment supplied by a local manufacturer and not by EEG machines manufactured internationally. Moreover, the LTMGH EEGs were recorded in a warm environment without air conditioning, which induces excessive delta power due to sweat artifacts (see Figure \ref{fig:boxplot_relative_power}). Consequently, the dataset could be prone to more artifacts, potentially increasing the challenges to detect abnormalities in the LTMGH EEGs reliably. As a result, we cannot train the EEG classifiers with this dataset unless we calibrate the system with this dataset. Moreover, we did not include segments from this dataset for the segment- and channel-level annotation to avoid confusion for the expert due to the abnormally high delta power.

We apply the following EEG preprocessing steps: a Butterworth notch filter ($4^{th}$ order) at 50Hz (Singapore and India) and 60Hz (USA), a 1Hz high-pass filter ($4^{th}$ order), and the Common Average Referential (CAR) montage. Next, we downsampled all the EEGs to 128Hz. Lastly, we applied artifact rejection based on noise statistics to remove high amplitude noise \cite{thomas2020automated}. This is achieved by computing the mean and standard deviation (std) of the root mean square (rms) amplitude of the EEG signal, then rejecting any 1s epoch (no overlap) with rms amplitude greater than $\text{mean} + 3 \times \text{std}$.

\subsection{EEG frequency features}
\label{feature}
We investigate the following EEG frequency bands: delta [1,4]Hz, theta [4,8]Hz, alpha [8,13]Hz, and beta [13,30]Hz. The relative power (RP) of each frequency band is calculated as:

\begin{equation} \label{eq:relativepower}
\text{RP}_{i} = \frac{\text{P}_{i}}{\text{P}_{\text{Total}}},
\end{equation}

\noindent where $\text{P}_{i}$ is the power in frequency band $i$, $\text{P}_{\text{Total}}=\sum{\text{P}_{i}}$, and $i \in [\delta, \theta, \alpha, \beta]$.

\begin{tablehere}
\vspace{-0.5 cm}
\centering
\tbl{Power ratios considered in the study. \label{tab:frequencyfeatures}}   
{
\scalebox{1.0}{
\begin{threeparttable}
\centering
\begin{tabular}{|c|c|} 
\hline
\textbf{Power Ratio} & \textbf{Definition} \\ 
\hline
\textbf{PRI} & ($\text{RP}_{\delta}$ + $\text{RP}_{\theta}$)/($\text{RP}_{\alpha}$ + $\text{RP}_{\beta}$)  \\
\textbf{DAR} & $\text{RP}_{\delta}$/$\text{RP}_{\alpha}$  \\
\textbf{TAR} & $\text{RP}_{\theta}$/$\text{RP}_{\alpha}$  \\
\textbf{TBAR} & $\text{RP}_{\theta}$/($\text{RP}_{\beta}$ + $\text{RP}_{\alpha}$)  \\
\hline
\end{tabular}

\begin{tablenotes}
\setlength\labelsep{0pt}
\footnotesize
\item
\end{tablenotes}

\end{threeparttable}
}}
\vspace{-0.5 cm}
\end{tablehere}

The RP is a normalized index as $\text{RP}_{\delta} + \text{RP}_{\theta} + \text{RP}_{\alpha} + \text{RP}_{\beta}=1$, and $\text{RP}_{i}$ $\in$ [0,1]. From the RP, we derive the power ratios (PR): Primary Ratio Index (PRI), Delta-Alpha-Ratio (DAR), Theta-Alpha-Ratio (TAR), and Theta-Beta-Alpha-Ratio (TBAR) (see Table \ref{tab:frequencyfeatures}). In this paper we consider the following eight spectral features: $\text{RP}_{\delta}$, $\text{RP}_{\theta}$, $\text{RP}_{\alpha}$, $\text{RP}_{\beta}$, PRI, DAR, TAR, and TBAR.

\subsection{Channel- and segment-level slowing annotation}
\label{annotation}
We have asked an expert to annotate individual channels and segments of EEGs from the TUH, NNI, Fortis, and NUH datasets (LTMGH dataset is omitted here). The channel-level annotations are required to train the channel-level detector in the SLDS and DLDS. We prepared 1000 5s EEG segments consisting of 900 unique segments and 100 duplicate segments (50 unique) for the expert to annotate on the channel- and segment-level.

\begin{tablehere}
\vspace{-0.4cm}
\tbl{Summary of annotated slowing segments and channels.\label{tab:segment_summary_basic}}
{
\centering
\scalebox{1.0}{
\centering
\begin{tabular}{|c|c|c|c} 
\cline{1-3}
\multicolumn{3}{|c|}{\textbf{Segment Annotation}} & \multicolumn{1}{l}{} \\ 
\cline{1-3}
 \textbf{Dataset}  & \textbf{Slow-free}  & \textbf{Slowing}  &  \\ 
\cline{1-3}
\textbf{TUH}  & 151 & 43 &  \\
\textbf{NNI}  & 142 & 94 &  \\
\textbf{Fortis}  & 169 & 65 &  \\
\textbf{NUH}  & 103 & 133 &  \\ 
\cline{1-3}
 \textbf{All}  & 565 & 335 &  \\ 
\cline{1-3}
\multicolumn{1}{c}{} & \multicolumn{1}{c}{} & \multicolumn{1}{c}{} &  \\ 
\hline
\multicolumn{4}{|c|}{\textbf{Channel Annotation}} \\ 
\hline
\textbf{Dataset} & \textbf{Slow-free} & \textbf{Slowing} & \multicolumn{1}{c|}{\textbf{Ambiguous}} \\ 
\hline
\textbf{TUH}  & 2869 & 263 & \multicolumn{1}{c|}{554} \\
\textbf{NNI}  & 2698 & 1473 & \multicolumn{1}{c|}{313} \\
\textbf{Fortis}  & 3211 & 1053 & \multicolumn{1}{c|}{182} \\
\textbf{NUH}  & 1957 & 1940 & \multicolumn{1}{c|}{587} \\ 
\hline
\textbf{All}  & 10735 & 4729 & \multicolumn{1}{c|}{1636} \\
\hline
\end{tabular}
}}
\vspace{-0.4cm}
\end{tablehere}

We select a segment duration of 5s, since the minimum cutoff frequency of a slow wave is 1Hz, corresponding to a period of 1s. Therefore, a 5s segment may contain up to five periods of slowing waveforms, hence it should be possible to detect slowing reliably. According to our findings, we choose the segments according to their PRI values, as it appears to be the most consistent for slowing classification. The annotations are performed by one expert in the NeuroBrowser (NB) software \cite{jing2016rapid}. We refer to the Appendix for more information on the annotation procedure. The number of slowing and slow-free segments and channels annotated from each dataset are displayed in Table \ref{tab:segment_summary_basic}. For the segment annotations, the segments are annotated as slow-free and slowing. 

We parsed the channel-level annotations as follows. For the segments labeled as slow-free, all channels are considered to be slow-free; indeed, if one of the channels exhibited slowing according to the annotator, he/she would have labeled the segment as `slowing'. Therefore, it is safe to treat all channels as slow-free. When the annotator labels a segment as `slowing', the annotator will also indicate at what channels `slowing' occurs. The remaining channels are not specifically marked as `slow-free'. They can contain other signal features such as artifacts, ictal activities, spikes, eye blinks, K-complexes, etc. They may even contain slowing, as the annotator may not have carefully annotated all channels. Therefore, we distinguish three cases: Slow-free channels (from slow-free segments), slowing channels (from slowing segments), as marked by the annotator, ambiguous channels (from slowing segments), not explicitly marked by the annotator.

\begin{tablehere}
\vspace{-0.4 cm}
\tbl{Summary of the PRI values extracted from segment annotations. \label{tab:unique_segment_summary}}   
{
\centering
\scalebox{0.95}{
\begin{tabular}{|c|c|c|c|c|c|c|} 
\hline
\multirow{2}{*}{\textbf{Dataset}} & \multicolumn{3}{c|}{\begin{tabular}[c]{@{}c@{}}\textbf{Slow-free}\\\textbf{Segment PRI} \end{tabular}} & \multicolumn{3}{c|}{\begin{tabular}[c]{@{}c@{}}\textbf{Slowing}\\\textbf{Segment PRI} \end{tabular}} \\ 
\cline{2-7}
 & \textbf{No}  & \textbf{Mean}  & \textbf{std}  & \textbf{No}  & \textbf{Mean}  & \textbf{std}  \\ 
\hline
\textbf{TUH}  & 151 & 5.544 & 3.366 & 43 & 15.647 & 11.721 \\ 
\hline
\textbf{NNI}  & 142 & 5.632 & 3.571 & 94 & 21.798 & 30.808 \\ 
\hline
\textbf{Fortis}  & 169 & 7.36 & 4.621 & 65 & 9.094 & 4.819 \\ 
\hline
\textbf{NUH}  & 103 & 7.409 & 5.986 & 133 & 17.722 & 13.213 \\ 
\hline\hline
 \textbf{All}  & 565 & 6.449 & 4.455 & 335 & 16.926 & 19.341 \\
\hline
\end{tabular}
}}
\vspace{-0.4 cm}
\end{tablehere}

\begin{table*}[t!]
\vspace{-0.4cm}
\tbl{Summary of the PRI values extracted from channel annotations. \label{tab:channel_summary}}
{
\centering
\scalebox{0.9}{
\centering
\begin{tabular}{|c|c|c|c|c|c|c|c|c|c|} 
\hline
\multirow{2}{*}{\textbf{Dataset}} & \multicolumn{3}{c|}{\begin{tabular}[c]{@{}c@{}}\textbf{Slow-free}\\\textbf{Channel PRI} \end{tabular}} & \multicolumn{3}{c|}{\begin{tabular}[c]{@{}c@{}}\textbf{Slowing}\\\textbf{Channel PRI} \end{tabular}} & \multicolumn{3}{c|}{\begin{tabular}[c]{@{}c@{}}\textbf{Ambiguous}\\\textbf{Channel PRI} \end{tabular}} \\ 
\cline{2-10}
 & \textbf{No}  & \textbf{Mean}  & \textbf{std}  & \textbf{No}  & \textbf{Mean}  & \textbf{std}  & \textbf{No}  & \textbf{Mean}  & \textbf{std}  \\ 
\hline
\textbf{TUH}  & 2869 & 5.544 & 5.185 & 263 & 25.192 & 33.307 & 554 & 11.115 & 11.189 \\ 
\hline
\textbf{NNI}  & 2698 & 5.632 & 7.668 & 1473 & 24.615 & 39.189 & 313 & 8.544 & 7.253 \\ 
\hline
\textbf{Fortis}  & 3211 & 7.360 & 8.492 & 1053 & 9.426 & 8.104 & 182 & 7.178 & 6.900 \\ 
\hline
\textbf{NUH}  & 1957 & 7.409 & 8.510 & 1940 & 18.573 & 18.689 & 587 & 14.913 & 17.218 \\ 
\hline \hline
\textbf{All}  & 10735 & 6.449 & 7.585 & 4729 & 18.786 & 27.025 & 1636 & 11.548 & 13.119 \\
\hline
\end{tabular}
}}
\vspace{-0.3 cm}
\end{table*}

\begin{figurehere}
\centering
\includegraphics[width=7cm]{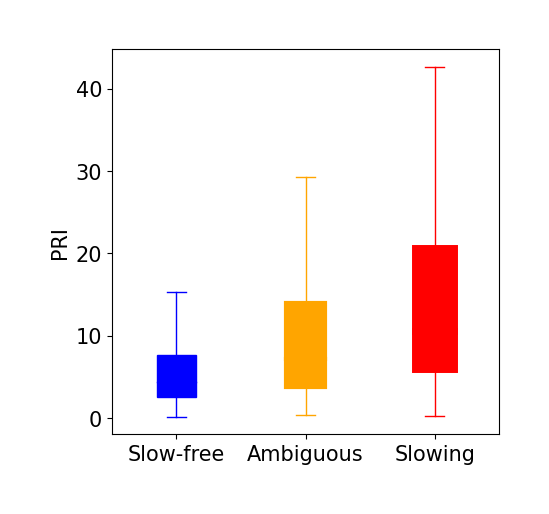}
\vspace{-0.6 cm}
\caption{PRI value distribution across slow-free, ambiguous, and slowing channels. Despite the fact that ambiguous channels are considered to contain no slowing in slowing segments, it has a relatively higher PRI value than slow-free channels, and lower PRI value than slowing channels.}
\label{fig:boxplot_PRI_channel}
\vspace{-0.2 cm}
\end{figurehere}

After obtaining the annotations, we removed the 100 duplicate segments (50 unique and 50 duplicates) from the 1000 segments and analyzed the PRI of the remaining 900 segments and channels in Table \ref{tab:unique_segment_summary}. The segments are split into two categories: slowing and slow-free. We notice that the PRI values of the Fortis segments have comparable PRI values for slowing and slow-free EEGs, which can lead to challenges during classification.

We summarize the distribution of the channel PRI values in Table \ref{tab:channel_summary} according to the three categories: slow-free, slowing, and ambiguous. We also display the boxplot of the PRI values across the channels from the slowing, slow-free, and ambiguous channels in Figure \ref{fig:boxplot_PRI_channel}. The ambiguous channels are supposed to be free of slowing but have higher PRI values than slow-free channels and lower PRI values than slowing channels. We discard the ambiguous channels from our analysis and training process to avoid false positives, as we cannot confidently assume they are slow-free. Again, the PRI values from the Fortis channels between slowing and slow-free are similar. In order not to confuse the classifiers, we decided to discard the ambiguous channels in the training data of the classifiers.

\begin{figure*}[htb]
\centering
\includegraphics[width=16.5cm]{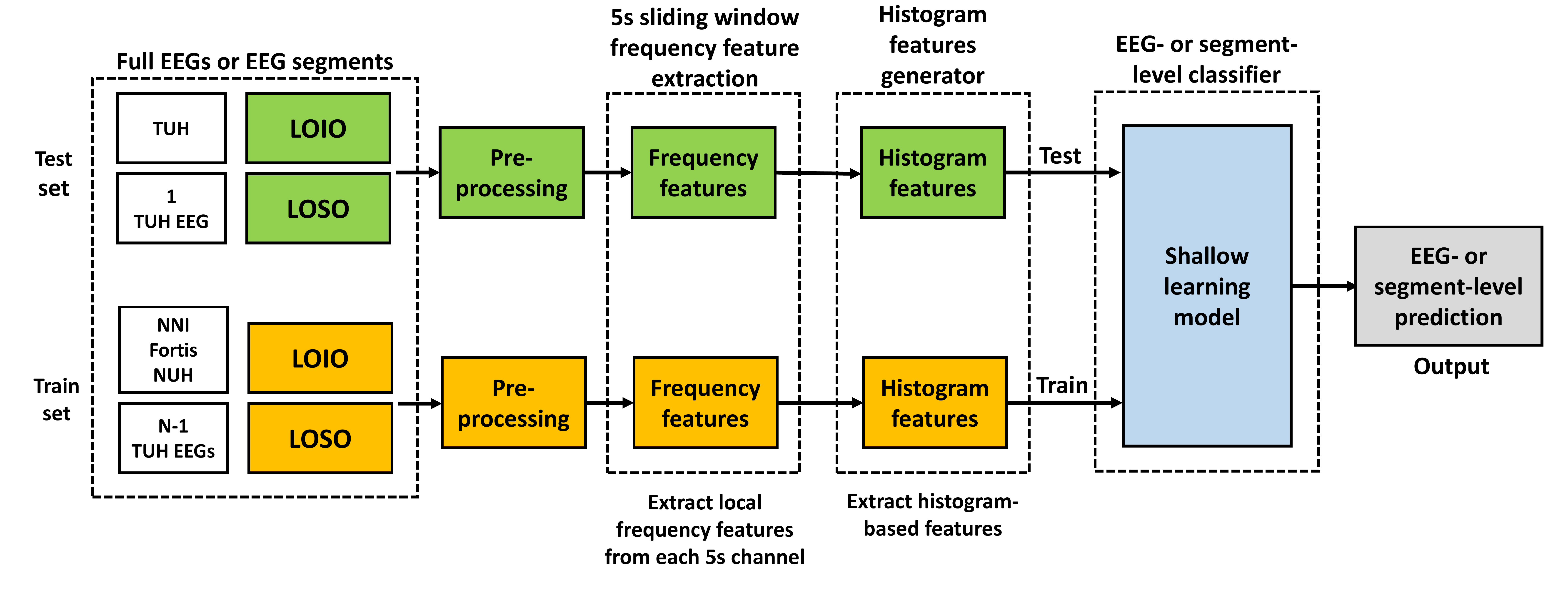}
\vspace{-0.4 cm}
\caption{LOIO and LOSO CV for the TDS for the segment- and EEG-level classification. Frequency features are extracted from single-channel segments and compiled into histograms and histogram-based features are extracted. A shallow learning model is trained to detect slowing in an EEG segment/full EEG from the histogram-based features. In LOIO CV, in each iteration, the dataset from one center is evaluated by a model trained on datasets from other centers. In LOSO CV, the EEG(s) of one subject is evaluated in each iteration by a model trained on the remaining EEGs from that same institution. LOSO CV is performed on each dataset independently. In the above, as an illustration, the model is tested on TUH data in LOIO and LOSO CV.}

\label{fig:EEG_classification_system_1}
\vspace{-0.2 cm}
\end{figure*}

\begin{figure*}[htb]
\centering
\includegraphics[width=16.5cm]{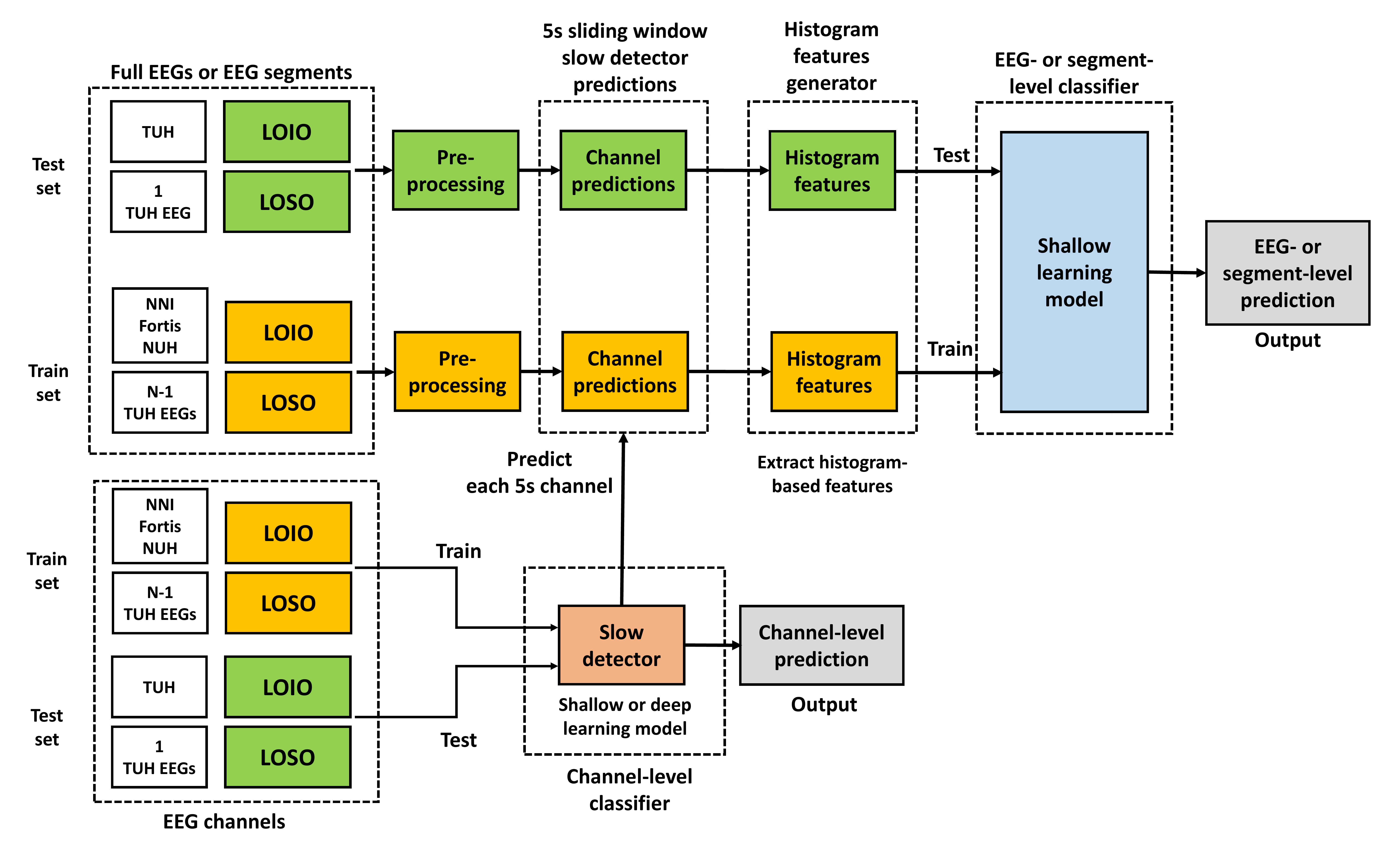}
\vspace{-0.4 cm}
\caption{LOIO and LOSO CV for the SLDS and DLDS for the channel-, segment-, and EEG-level classification. We train the channel-level slowing detector (shallow or deep learning model) on the channel-wise annotated EEG dataset. We arrange the channel-level detector outputs into histograms and extract features from those histograms for segment-level and EEG-level slowing detection. Next, we train a shallow learning model with the histogram-based features to detect slowing in EEG segment/full EEG. In the above as an illustration, the model is tested on TUH data in LOIO and LOSO CV.}
\label{fig:EEG_classification_system_2}
\vspace{-0.2 cm}
\end{figure*}


\subsection{Proposed EEG classification systems}
\label{classification_systems}
This section outlines the three proposed EEG classification systems: TDS, SLDS, and DLDS. We evaluate the systems on channel-, segment-, and EEG-level slowing detection. The systems pipelines are summarized in Table \ref{tab:EEG_system_summary}. The TDS for the segment- and EEG-level classification is illustrated in Figure \ref{fig:EEG_classification_system_1}, while the SLDS and DLDS for the channel-, segment-, and EEG-level classification are illustrated in Figure \ref{fig:EEG_classification_system_2}. The classification systems are named after the channel-level detector; the TDS, SLDS, and DLDS rely on thresholding, shallow learning, and deep learning channel-level classifiers.

\begin{tablehere}
\vspace{-0.4cm}
\tbl{Parameters of the shallow learning models.\label{tab:shallow_learning_parameters}}
{
\centering
\scalebox{0.85}{
\centering
\begin{tabular}{|c|c|c|} 
\hline
\textbf{Shallow Learning Model} & \textbf{Parameter} & \textbf{Value} \\ 
\hline
\multirow{2}{*}{\begin{tabular}[c]{@{}c@{}}\textbf{Logistic}\\\textbf{Regression}\end{tabular}} & Solver & lbfgs \\
 & Max iteration & 10000000 \\ 
\hline
\multirow{3}{*}{\textbf{SVM}} & Kernel & \{linear, rbf\} \\
 & C & 1 \\
 & Gamma & scale \\ 
\hline
\multirow{2}{*}{\begin{tabular}[c]{@{}c@{}}\textbf{Gradient}\\\textbf{Boosting}\end{tabular}} & Estimators & 100 \\
 & Max features & 1 \\ 
\hline
\textbf{AdaBoost} & Estimators & 100 \\ 
\hline
\multirow{3}{*}{\begin{tabular}[c]{@{}c@{}}\textbf{Random}\\\textbf{Forest}\end{tabular}} & Max depth & 4 \\
 & Estimators & 100 \\
 & Max features & 1 \\
\hline
\end{tabular}
}}
\vspace{-0.4cm}
\end{tablehere}

Additionally, all three systems deploy a shallow learning model to perform segment- and EEG-level prediction. We applied five different shallow learning classifiers in this study: {Logistic Regression (LR)\cite{raschka2015python}}, {SVM (linear and Gaussian/rbf kernel)\cite{smola2004tutorial}}, {Gradient Boosting (GB)\cite{friedman2002stochastic}}, {AdaBoost\cite{freund1995desicion}}, and {Random Forest (RF)\cite{liaw2002classification}}. The parameters of the shallow learning models are summarized in Table \ref{tab:shallow_learning_parameters}. We apply three feature processing steps to the training data to enable efficient training. First, we apply a threshold to the standard deviation (std) to remove non-significant features (std $\leq10^{-7}$). Then, we standardize the features by subtracting the mean and dividing by the std. Lastly, we apply Synthetic Minority Over-sampling Technique (SMOTE) with five nearest neighbors to construct synthetic samples of the minority class to match the majority class for training \cite{chawla2002smote}. By applying SMOTE to re-balance any imbalanced dataset, we reduce classification bias and improve classification accuracy.


\vspace{-0.5 cm}
\subsubsection{Channel-level classification}
Here we describe the single-channel segment (channel-level) slowing detection in the TDS, SLDS, and DLDS. The channel-level slowing detector allows us to identify differences between slowing EEG waveforms and background EEG signals. The channel-level detector is important as it can be deployed to detect slowing in individual channels and time instances in an EEG, thereby allowing us to determine where and when the slowing abnormality occurs.

In the TDS, we apply simple thresholding on the eight spectral features computed at single-channel segments. If the spectral feature is above (for $\text{RP}_{\delta}$, $\text{RP}_{\theta}$, PRI, DAR, TAR, and TBAR) or below (for $\text{RP}_{\alpha}$ and $\text{RP}_{\beta}$) the threshold, the waveform at that channel exhibits slowing. For the SLDS, we train a shallow learning model on the eight spectral features to detect EEG slowing at the channel-level. Finally, for the DLDS, we train a CNN, whose input is the spectrum of the EEG waveforms.

The EEG spectrum is obtained by transforming the 5s EEG signal (640 samples) to the frequency domain [0,64]Hz (321 samples). We discard the frequencies in the [30,64]Hz band to eliminate the gamma band component, keeping only the [0,30]Hz band (150 samples). Finally, we smoothen the spectrum with a moving average (length five). The input of the CNN is the smoothed spectrum (150 samples). We implemented the CNN in Keras 2.2.0~\cite{gulli2017deep} on an Nvidia GeForce GTX 1080 Graphical Processing Unit (GPU) with Ubuntu 16.04 as the operating system. 

\begin{figure*}
\centering
\includegraphics[width=16cm]{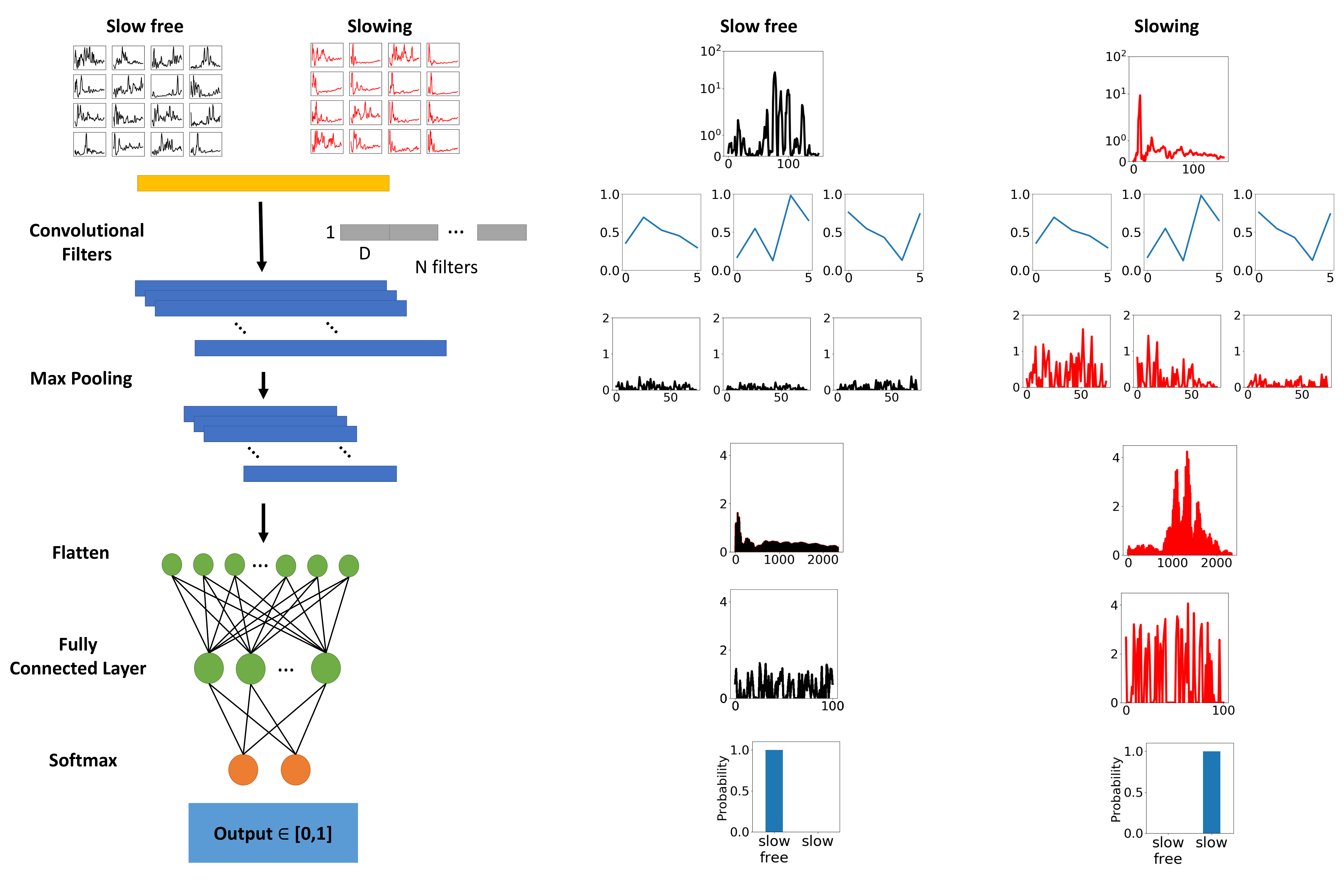}
\vspace{-0.4 cm}
\caption{1D CNN architecture adopted in the study (left), and the activations along with each layer of the CNN for a slow-free (middle) and slowing EEG single-channel segment (right). The input of the CNN is the 5s single-channel normalized frequency spectrum. The filters of the convolutional layer of the CNN and their corresponding activations are displayed, together with the activations in the fully connected layer. The results show a distinct difference between slow-free and slowing signals, which the CNN learned to perform slowing predictions on channel-level.}
\label{fig:CNN_model}
\vspace{-0.1 cm}
\end{figure*}

We devise the CNN detector (see Figure \ref{fig:CNN_model}) comparable to the 1D CNN architecture proposed by Thomas et al. \cite{thomas2018eeg}. First, the convolutional operation is performed by convolving the smoothened EEG spectrum with optimized 1D convolution filters. Next, the resulting convolution outputs are passed through non-linear activation functions. Specifically, we chose Rectified Linear Units (ReLU) as the activation functions. The outputs of these activation functions together form spectral feature maps. The dimensions of the feature maps are reduced by max-pooling. Next, the features are flattened and fed into a fully connected layer. The fully connected layer outputs are mapped into [0,1] with a softmax function, where 0 and 1 correspond to a slow-free and slowing EEG signal, respectively. We decided to include few layers and short filters in the CNN in order to limit the number of trained parameters, and consequently, to reduce the computation time.

We organized the training samples in mini-batches whose size is equal to half the number of slowing waveforms in the training set. To prevent overfitting, we applied balanced training by generating mini-batches with the same number of randomly selected slow waveforms and normal (background) waveforms. Additionally, a dropout of 0.5 is applied in the fully connected layer, a common choice in the literature.

\vspace{-0.4 cm}
\begin{tablehere}
\tbl{Optimized hyperparameters in the CNN. \label{tab:CNN_details}}
{
\centering
\scalebox{0.95}{
\begin{threeparttable}

\begin{tabular}{|c|c|} 
\hline
 \textbf{Parameters}  & \textbf{Values/Type}  \\ 
\hline
\textbf{Number of convolution layers}  & 1, 2, 3 \\ 
\hline
\textbf{Number of fully connected layers}  & 1, 2, 3 \\ 
\hline
\textbf{Number of convolution filters}  & \begin{tabular}[c]{@{}c@{}}8, 16, 32, 64, 128 \end{tabular} \\ 
\hline
\begin{tabular}[c]{@{}c@{}}\textbf{Dimension of convolution filters}\\\textbf{(kernel)} \end{tabular} & \begin{tabular}[c]{@{}c@{}}1$\times$3, 1$\times$5, 1$\times$7,\\1$\times$9, 1$\times$11, 1$\times$13 \end{tabular} \\ 
\hline
\textbf{Number of hidden neurons}  & 100 \\ 
\hline
\textbf{Activation functions}  & ReLU \\ 
\hline
\textbf{Dropout probability}  & 0.5 \\ 
\hline
\textbf{Size of the batch processing}  & $\frac{n_{s}}{2}$ \\
\hline
\textbf{Maximum number of iterations}  & 10000 \\ 
\hline
\textbf{Optimizer}  & Adam \\ 
\hline
\textbf{Learning rate}  & 10E-4 \\ 
\hline
\textbf{Measure}  & Cross-entropy \\
\hline
\end{tabular}

\begin{tablenotes}
\setlength\labelsep{0pt}
\footnotesize
\item $n_{s}$: Total number of annotated waveforms.
\end{tablenotes}

\end{threeparttable}
}}
\vspace{-0.5 cm}
\end{tablehere}

The hyperparameters of the CNN are optimized by applying a nested CV on the training data: 80\% of the training data is utilized for learning the classifier parameters; the rest is used for validation, i.e., for selecting the CNN hyperparameters and for deciding when to stop the training process \cite{thomas2020automated}. The CNN training is halted when the validation cost reaches its minimum. Table \ref{tab:CNN_details} lists the settings of the hyperparameters evaluated in our tests. We chose cross-entropy as the objective function of the CNN, and we optimized it by the Adam algorithm \cite{kingma2014adam}.

\subsubsection{Segment- and EEG-level classification}
All three systems detect slowing at the segment-level and EEG-level by exploiting statistics computed from the individual channels. While detecting slowing in individual channels is critical, it cannot explicity inform us of any abnormality in the EEG itself, as a single time instance or channel of slowing is insufficient evidence that an anomaly had occurred. For EEG classification, we need slowing information across the entire EEG. Therefore, we rely on histogram-based features for segment- and EEG-level classification.

To detect slowing in a 5s 19-channel EEG segment, we extract statistics from the 19 channels and then arrange those 19 values into a histogram. For the TDS and SLDS/DLDS, the statistics are a selected spectral feature and output of the channel-level slowing detector, respectively. Similarly, when detecting slowing in a full EEG, we first split the full EEG into $n$ 5s segments with a 75\% overlap. Next, we extract statistics from all those segments and arrange those 19$n$ values into a histogram. Also, in this case, the statistics are a selected spectral feature and output of the channel-level slowing detector for the TDS and SLDS/DLDS, respectively. We tested several values for the number of bins: 2, 5, 10, 15, or 20 bins. We have reported the selected number of bins in Table \ref{tab:LOIO} and \ref{tab:LOSO}, leading to the highest BAC. We conducted both segment- and EEG-level classification to illustrate the flexibility of our systems, capable of performing slowing detection for short 5s segments to routine EEGs of 30 minutes and beyond.

For the TDS, we select one of the eight spectral features ($\text{RP}_{\delta}$, $\text{RP}_{\theta}$, $\text{RP}_{\alpha}$, $\text{RP}_{\beta}$, PRI, DAR, TAR, and TBAR) to form the histogram. As different spectral features have different ranges of values for slowing and slow-free EEGs, we must normalize those features extracted across all the single-channel segments. To do so, for each dataset, we randomly select 50 slow-free EEG and compute the histogram of the selected spectral feature, and find the value at $\text{mean} + 3 \times \text{std}$. We perform min-max normalization by dividing that value to all features extracted from the single-channel segments to ensure that most of the values in slow-free EEGs are bounded between approximately [0,1]. 

To include the slowing portions exceeding the range of [0,1] (PR for slowing EEG is always greater than in slow-free EEG), we increase the range to [0,4] (see Figure \ref{fig:histogram_uls}). Additionally, we include two additional bins at [-100,0) and (4, 100] to include the outliers but not significantly skew the histogram distribution.

\begin{figurehere}
\centering
\includegraphics[width=8cm]{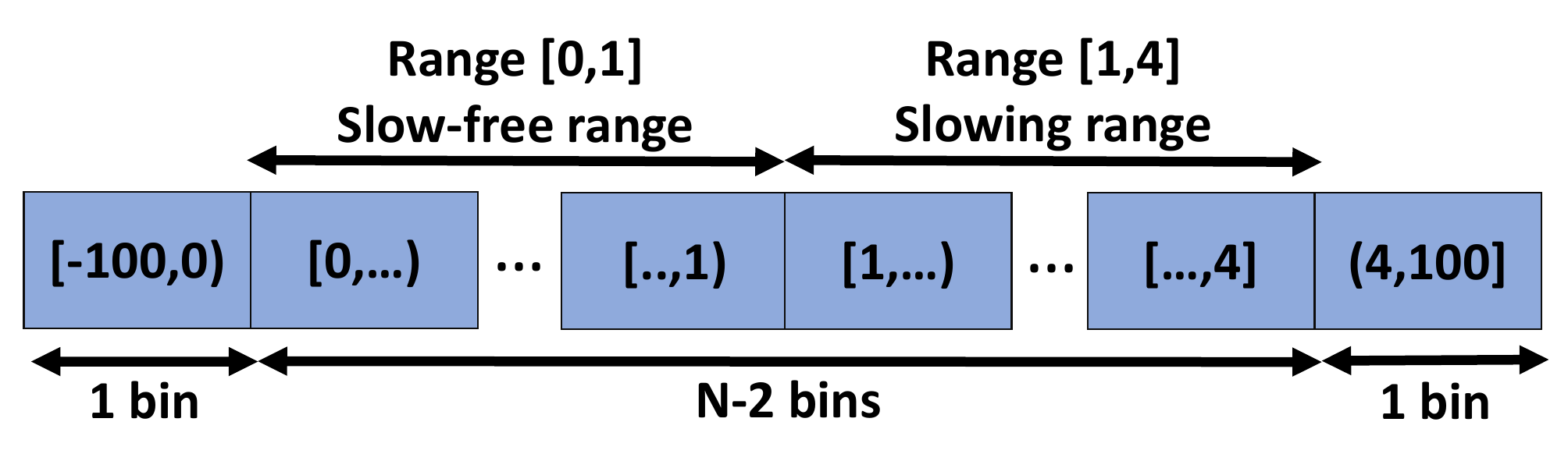}
\vspace{-0.8 cm}
\caption{Histogram of TDS.}
\label{fig:histogram_uls}
\vspace{-0.1 cm}
\end{figurehere}

On the other hand, for the SLDS and DLDS, the selected feature is the output of the single-channel slowing detector (bounded to [0,1]). Hence, we do not need to determine the upper and lower limit of the slowing probability, making the computation process much easier. We apply the single-channel slowing detector on all the channels in the 5s EEG segments and arrange the outputs into a histogram. 

From the histograms, we extract the following features: histogram counts, mean, median, mode, standard deviation, minimum value, maximum value, range, kurtosis, and skewness of the histogram. Histogram features are easy to interpret and are straightforward to handle numerically. Finally, all three systems deploy a shallow learning model that takes the histogram-based features as input to perform a segment- or EEG-level prediction.

\begin{figure*}
\vspace{-0.5 cm}
\begin{minipage}[b]{0.24\linewidth}
  \centering
  \centerline{\includegraphics[width=4.4cm]{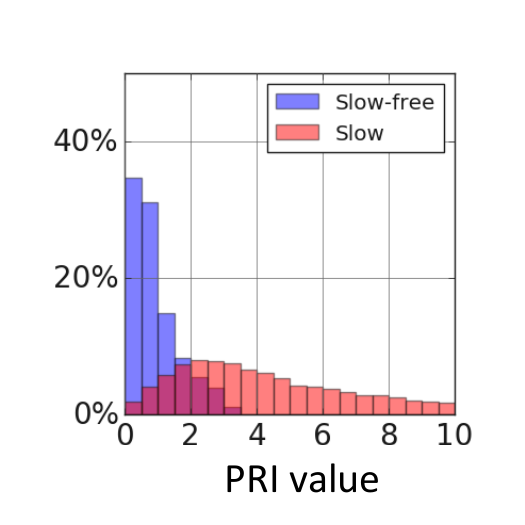}}
  \vspace{-0.3 cm}
  \centerline{(a) TUH PRI histogram.}\medskip
\end{minipage}
\hfill
\begin{minipage}[b]{0.24\linewidth}
  \centering
  \centerline{\includegraphics[width=4.4cm]{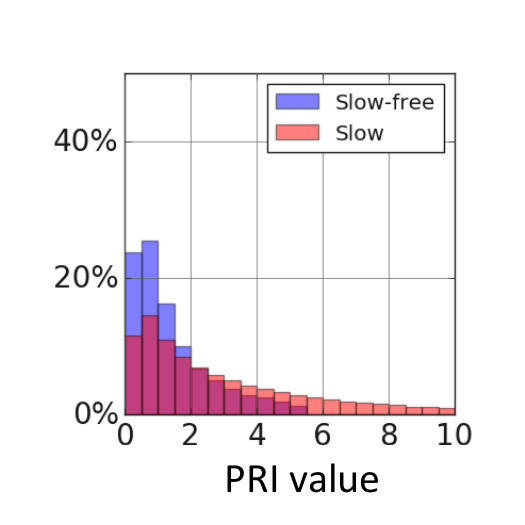}}
  \vspace{-0.3 cm}
  \centerline{(b) NNI PRI histogram.}\medskip
\end{minipage}
\hfill
\begin{minipage}[b]{0.24\linewidth}
  \centering
  \centerline{\includegraphics[width=4.4cm]{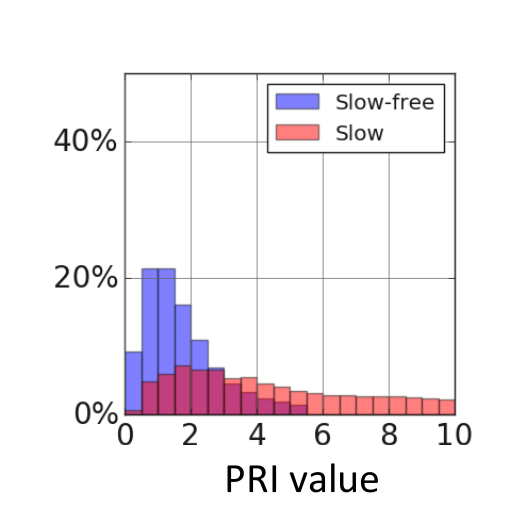}}
  \vspace{-0.3 cm}
  \centerline{(c) Fortis PRI histogram.}\medskip
\end{minipage}
\hfill
\begin{minipage}[b]{0.24\linewidth}
  \centering
  \centerline{\includegraphics[width=4.4cm]{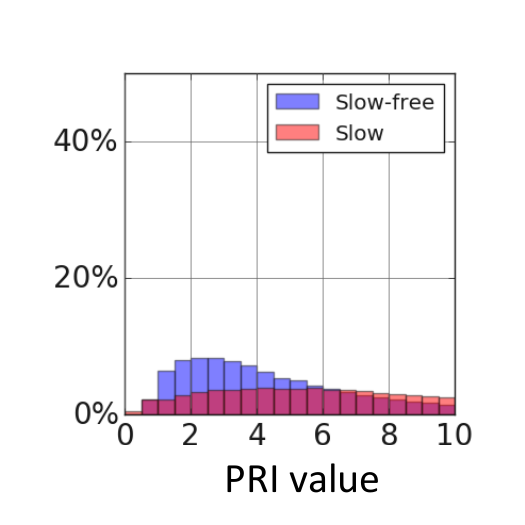}}
  \vspace{-0.3 cm}
  \centerline{(d) LTMGH PRI histogram.}\medskip
\end{minipage}
\vspace{-0.4 cm}
\caption{Histograms of PRI values computed from 5s EEG segments. a) TUH, b) NNI, c) Fortis, d) LTMGH. Slowing EEGs have a higher mean and wider distribution of PRI than slow-free EEGs. Additionally, slow-free EEGs can contain high PRI segments, but less frequently. For LTMGH EEGs, the PRI distribution is less distinct, where even slow-free EEGs can contain a substantial number of high PRI segments.}
\label{fig:histogram_PRI}
\vspace{-0.3 cm}
\end{figure*}

To understand why the histogram-based features are suitable for classification, we display PRI histograms for slowing and slow-free EEGs in Figure \ref{fig:histogram_PRI}. While slow-free EEGs have a lower average PRI, they still contain a small percentage of single-channel segments with high PRI values. As high PRI values are associated with slowing, this suggests that even slow-free EEGs can contain some amount of abnormal slowing, but less frequently than a slowing EEG. Consequently, the histogram-based features can account for the slowing frequency distribution in EEG segments and full EEGs and can, therefore, serve as useful input features for machine learning methods for detecting slowing in EEG.

\subsection{Datasets for training and testing}
We conduct both LOIO and LOSO CV for evaluating the proposed slowing detection systems. In Table \ref{tab:dataset_LOIO_LOSO}, we list the various datasets involved in the training and testing of the detection systems in LOIO and LOSO CV. The NUH dataset is always included for training the channel-level slowing detector for EEG-level LOIO CV, as we do not perform EEG-level classification on full EEGs from the NUH dataset. This is because we do not have the slowing labels for the full EEGs for the NUH dataset. The LTMGH dataset is always excluded in the training processes during EEG-level LOIO CV. It is only deployed when evaluating the LTMGH dataset itself with LOSO CV. As we do not have annotated channels from the LTMGH dataset, we cannot perform an EEG-level LOSO CV on the LTMGH dataset with the SLDS and DLDS. Instead, we perform a modified LOSO CV, by training the channel-wise detector with channel data from other datasets as a substitute.

\begin{table*}[t!]
\centering
\tbl{Datasets allocation during training and testing for LOIO and LOSO CV.\label{tab:dataset_LOIO_LOSO}}
{
\scalebox{1}{
\begin{threeparttable}
\begin{tabular}{|c|c|c|c|c|c|c|} 
\hline
\multirow{4}{*}{\textbf{Testing set}} & \multicolumn{6}{c|}{\textbf{Training set}} \\ 
\cline{2-7}
 & \multicolumn{2}{c|}{\textbf{Channel/Segment-level}} & \multicolumn{4}{c|}{\textbf{EEG-level}} \\ 
\cline{2-7}
 & \multirow{2}{*}{\textbf{LOIO}} & \multirow{2}{*}{\textbf{LOSO}} & \multicolumn{2}{c|}{\textbf{LOIO}} & \multicolumn{2}{c|}{\textbf{LOSO}} \\ 
\cline{4-7}
 &  &  & \textbf{Channel-level} & \textbf{EEG-level} & \textbf{Channel-level} & \textbf{EEG-level} \\ 
\hline
\textbf{TUH} & NNI, Fortis, NUH & TUH & NNI, Fortis, NUH & NNI, Fortis & TUH & TUH \\ 
\hline
\textbf{NNI} & TUH, Fortis, NUH & NNI & TUH, Fortis, NUH & TUH, Fortis & NNI & NNI \\ 
\hline
\textbf{Fortis} & TUH, NNI, NUH & Fortis & TUH, NNI, NUH & TUH, NNI & Fortis & Fortis \\ 
\hline
\textbf{NUH} & TUH, NNI, Fortis & NUH & - & - & - & - \\ 
\hline
\textbf{LTMGH} & - & - & \begin{tabular}[c]{@{}c@{}}TUH, NNI,\\Fortis, NUH\end{tabular} & \begin{tabular}[c]{@{}c@{}}TUH, NNI,\\Fortis\end{tabular} & \begin{tabular}[c]{@{}c@{}}TUH, NNI,\\Fortis, NUH\end{tabular} & LTMGH \\
\hline
\end{tabular}

\begin{tablenotes}
\setlength\labelsep{0pt}
\footnotesize
\item 
\end{tablenotes}

\end{threeparttable}
}}
\vspace{-0.6 cm}
\end{table*}

\section{Results}
\label{results}

\subsection{EEG relative power}

\begin{figure*}
\begin{minipage}[b]{0.24\linewidth}
  \centering
  \centerline{\includegraphics[width=4cm]{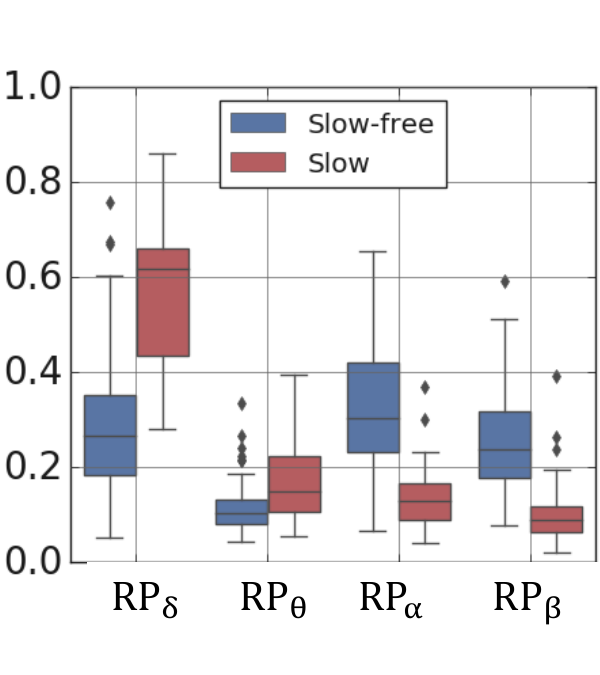}}
  \vspace{-0.3 cm}
  \centerline{(a) TUH RP.}\medskip
\end{minipage}
\hfill
\begin{minipage}[b]{0.24\linewidth}
  \centering
  \centerline{\includegraphics[width=4cm]{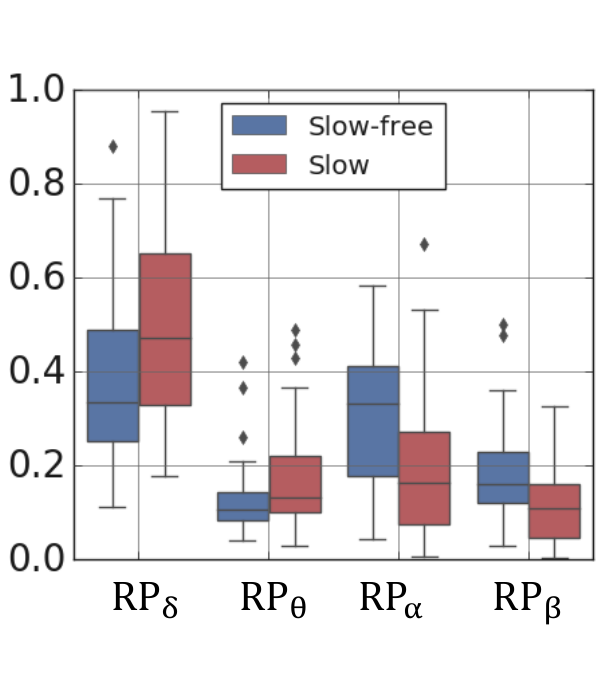}}
  \vspace{-0.3 cm}
  \centerline{(b) NNI RP.}\medskip
\end{minipage}
\hfill
\vspace{-0.2 cm}
\begin{minipage}[b]{0.24\linewidth}
  \centering
  \centerline{\includegraphics[width=4cm]{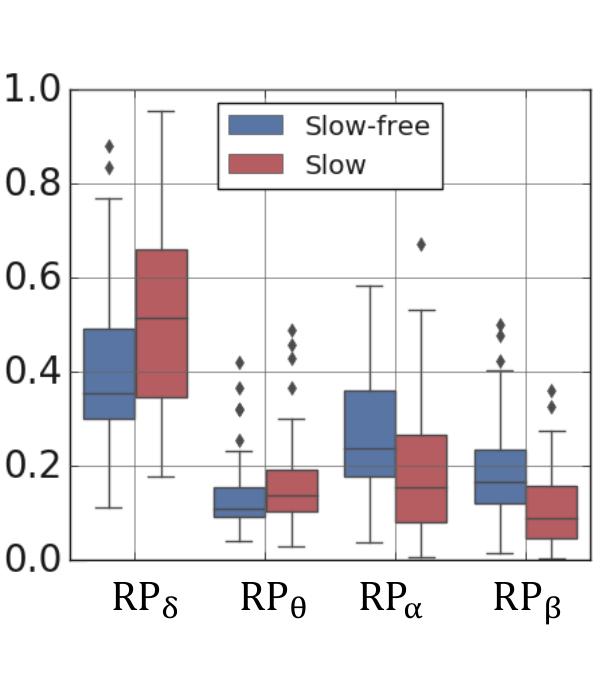}}
  \vspace{-0.3 cm}
  \centerline{(c) Fortis RP.}\medskip
\end{minipage}
\hfill
\begin{minipage}[b]{0.24\linewidth}
  \centering
  \centerline{\includegraphics[width=4cm]{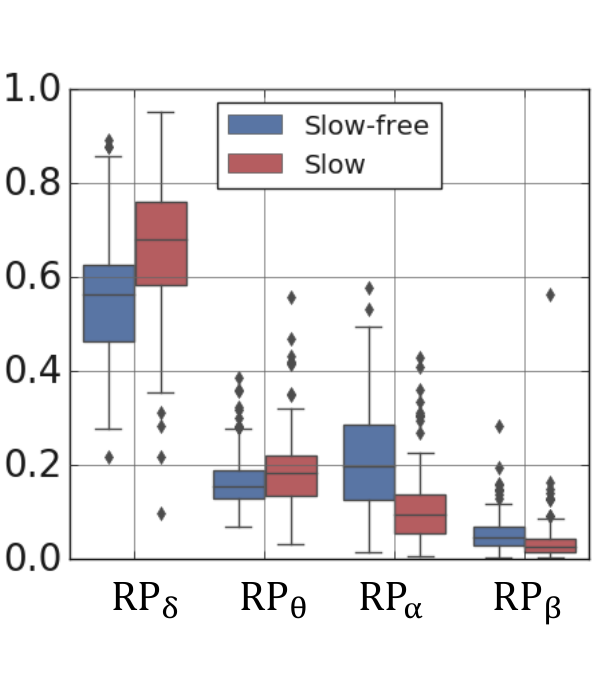}}
  \vspace{-0.3 cm}
  \centerline{(d) LTMGH RP.}\medskip
\end{minipage}
\vspace{-0.1 cm}
\caption{Relative power in the delta, theta, alpha, and beta band: a) TUH, b) NNI, c) Fortis, d) LTMGH. The delta and theta power is stronger in EEGs that exhibit slowing compared to slow-free EEGs. The delta power in LTMGH EEGs is significantly higher than in EEGs from the other datasets.}
\label{fig:boxplot_relative_power}
\vspace{-0.3 cm}
\end{figure*}

We compare the relative power (RP) of EEGs of the datasets in Figure \ref{fig:boxplot_relative_power}. The NUH dataset is not included due to the lack of slowing labels. Slowing EEGs have a higher delta and theta RP, with lower alpha and beta RP than slow-free EEGs. The RP values in the EEGs from the LTMGH dataset are significantly different from those from the TUH, NNI, and Fortis datasets. The EEGs from LTMGH have higher delta RP and a much smaller beta RP. Therefore, it is more meaningful to analyze the LTMGH dataset separately.

\subsection{Intra-rater agreement (IRA)}
In this section, we address the label intra-rater agreement (IRA) of the expert. In this case, the IRA is the percentage of agreement of the labels between the duplicated segments. Theoretically, the slowing detector from the SLDS and DLDS cannot outperform the IRA of the expert, for they are trained with the annotations from the expert. The IRA gives us an approximate upper-limit on the accuracy of our proposed systems. 

The channel- and segment-level IRA is 72.4\% and 82\%, respectively. The disagreements are mainly due to artifacts, eye blinks, or interictal epileptiform discharges (IEDs), matching observation with the literature \cite{grant2014eeg}. Also, a study performed by Piccinelli et al. reported an IRA of 88.6\% for expert agreement for classifying EEGs into three classes: EEGs with IEDs, EEGs with slow waves, and normal EEGs \cite{piccinelli2005inter}. Another study on the IRA of IEDs in EEG reported that the median IRA between 9 experts is 80\%, comparable to our current observation \cite{noe2020most}.

\subsection{Classification results}
The best results for the channel-, segment-, and EEG-level LOIO and LOSO CV for each system, together with their parameters, are displayed in Table \ref{tab:LOIO} and \ref{tab:LOSO}. We deployed the following performance measures: area under the receiver operating characteristic curve (AUC), balanced accuracy (BAC), sensitivity (SEN), and specificity (SPE). Since the number of slowing and slow-free cases is sometimes imbalanced for all three classification tasks, we evaluate the results mainly in terms of BAC.

\begin{equation} \label{eq:SEN}
\text{SEN} = \frac{\text{TP}}{\text{TP}+\text{FN}},
\end{equation}

\begin{equation} \label{eq:SPE}
\text{SPE} = \frac{\text{TN}}{\text{TN}+\text{FP}},
\end{equation}

\begin{equation} \label{eq:BAC}
\begin{split}
\text{BAC} & = \frac{1}{2} \times \left( \frac{\text{TP}}{\text{TP}+\text{FN}} + \frac{\text{TN}}{\text{TN}+\text{FP}} \right) \\
& = \frac{1}{2} \times (\text{SEN} + \text{SPE}),
\end{split}
\end{equation}

\noindent where $\text{TP}$, $\text{TN}$, $\text{FP}$, and $\text{FN}$ is the true positive, true negative, false positive, and false negative of the classification results, respectively.

\subsubsection{Channel-level classification results}
For LOIO CV, the SLDS and DLDS yield the best performance, both achieving a mean BAC of 71.9\%. The TDS obtains a mean BAC of 68.4\%. For LOSO CV, the DLDS system yields an impressive mean BAC of 72.4\%, besting both the TDS and SLDS. The DLDS performed the best for both cases.

The TDS that deploys thresholding on the PRI achieved the best LOIO and LOSO CV mean BAC, suggesting that PRI is the optimal feature for channel-level slowing identification. All three systems did not perform well on the Fortis dataset while achieving the best mean BAC on the NNI dataset. The LOIO and LOSO CV results from the three systems achieved comparable channel-level classification accuracy to the channel-level IRA of the expert of 72.4\%. 

To understand what is learned in the CNN slowing detector in the DLDS, we analyze the feature maps of the convolutional layer in the CNN, as shown in Figure \ref{fig:CNN_model}. The feature maps revealed that the convolution layer assigns weights in a seemingly random manner to different frequencies in the spectrum (see Figure \ref{fig:filter}). These optimized quasi-random 1D convolution filters are similar to purely random convolution filters, which were commonly applied in the past to avoid learning the CNN filters, and have been shown to perform well even with limited data \cite{saxe2011random, saxe2013exact}. 

\begin{figurehere}
\centering
\includegraphics[width=8cm]{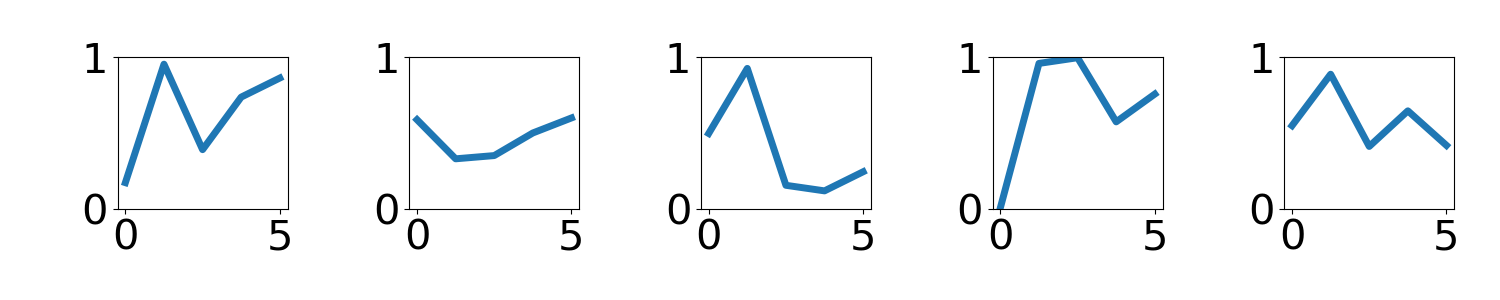}
\vspace{-0.6cm}
\caption{Sample filters with filter length of 5 deployed by the CNN. The filters are optimized by the CNN, but can appear random.}
\label{fig:filter}
\vspace{-0.5cm}
\end{figurehere}

\vspace{0.1 cm}
\begin{figurehere}
\centering
\includegraphics[width=5.0cm]{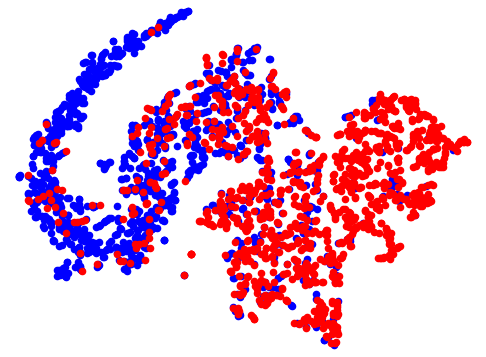}
\vspace{-0.1 cm}
\caption{Two-dimensional embedding of the 100-dimensional second fully connected layer in the CNN obtained by t-SNE. The slow-free and slowing EEG segments are marked in blue and red, respectively.}
\label{fig:t-SNE}
\vspace{-0.3 cm}
\end{figurehere}

As a verification, we mapped the second fully connected layer of the CNN into a 2D plane through t-Distributed Stochastic Neighbour Embedding (t-SNE) (see Figure \ref{fig:t-SNE}) \cite{maaten2008visualizing}. We can observe separable clusters, indicating that the neurons in the fully connected layers learn a meaningful representation of the EEG waveforms. 

\subsubsection{Segment-level classification results}
For LOIO CV, the DLDS achieves the best mean BAC, followed by the SLDS and the TDS. On the other hand, the DLDS yields the best system performance for the LOSO CV, followed by the TDS and the SLDS. The DLDS yields the best performance for both LOIO and LOSO CV, achieving a mean BAC of 75.5\% and 76.6\%, respectively. This BAC is close to the segment IRA of the expert of 82\%. Similarly, employing PRI to construct the histograms yielded the best LOIO and LOSO CV results for the TDS. All three systems performed the worst on the Fortis dataset while achieving the best BAC on the NNI dataset.

\subsubsection{EEG-level classification results}
For LOSO CV, we present the results for classification both with and without the LTMGH dataset. We applied LOSO CV on the LTMGH EEGs to verify whether the proposed systems perform well on those EEGs after recalibration of the EEG-level classifiers. Since the LTMGH EEGs appears to be less generalizable, we also report the average results for LOSO CV, excluding those EEGs. For the same reason, we also exclude those EEGs from the LOIO CV analysis.

For LOIO CV (excluding LTMGH dataset), the DLDS achieved the highest mean BAC of 82.0\%, with the TDS and SLDS reaching a mean BAC of 80.6\% and 79.7\%, respectively. However, if we include the LTMGH dataset, the TDS obtains the best mean BAC of 75.7\%, which is substantially lower, as the system performed poorly on the LTMGH dataset. All three systems lead to unsatisfactory classification results for LOIO CV on the LTMGH dataset, as the LTMGH EEGs do not match well with the EEGs from other datasets. Therefore, for EEGs that are not generalizable (e.g., enhanced delta power as in the LTMGH EEGs), we must recalibrate the EEG-level classification systems. To assess the improvement after recalibration, we perform LOSO CV on all datasets, including LTMGH EEGs.

For LOSO CV (including the LTMGH dataset), the TDS achieved the best mean BAC of 80.5\%, with a decent classification BAC of 75.8\% on the LTMGH dataset. Meanwhile, the SLDS and DLDS achieved a mean BAC below 80.0\%, with a BAC of around 70.0\% on the LTMGH dataset. 

One of the reasons that the SLDS and DLDS yield poor results for the LTMGH EEGs can be because we performed a modified LOSO CV on the LTMGH dataset (we do not have labeled channel data from LTMGH to train the channel-level classifier). Hence, the SLDS and DLDS may not be able to detect channel-wise slowing accurately. Instead, deploying a classification system without a channel-level detector such as the TDS to perform EEG-level classification may resolve this issue. 

When we exclude the LTMGH dataset, all three systems yield an approximately identical LOSO CV mean BAC of 82.0\%. This observation implies that the three systems can generate the same EEG-level classification accuracy after recalibration with EEGs from a particular dataset, despite the different system pipelines. However, this only applies to EEGs recorded under standard conditions (EEGs from TUH, NNI, and Fortis).  

If we do not have access to EEG reports to recalibrate the EEG classifiers, the LOIO CV results suggest that the systems could evaluate the EEGs as reliably as a recalibrated system. Omitting the LTMGH dataset, the three systems achieved an LOIO CV mean BAC close to the LOSO CV mean BAC of 82.0\% achieved by all three systems.

The DLDS achieves an almost identical mean BAC of approximately 82.0\% for both LOIO and LOSO CV (excluding the LTMGH dataset). This implies that the DLDS can potentially perform equally well in both scenarios. Moreover, this is the best BAC that we have obtained for the given current datasets and clinical reports.

In summary, we have demonstrated the need to evaluate the systems via both LOIO and LOSO CV. The LOSO CV BAC values are usually better than the ones in LOIO CV, since the EEG classifiers are trained and tested on EEGs of the same institutions. Therefore, the classifiers are effectively recalibrated to the EEGs of that institution. Hence, if EEG data is available, it is advisable to retrain the EEG classifiers on EEG data (and corresponding reports) from the institution where it will be deployed.

If such data is unavailable, our LOIO CV results suggest that reliable detection of EEG slowing can still be achieved through EEG classifiers trained on EEGs from other institutions. For EEG that do not generalize well, retraining of the EEG classifiers might be required; we have shown for the LTMGH EEGs that reliable slowing detection can be obtained after recalibration.

\begin{table*}[tp!]
\centering
\tbl{Channel-, segment- and EEG-level LOIO CV results for the different datasets.\label{tab:LOIO}}
{
\scalebox{1}{
\begin{threeparttable}

\begin{tabular}{|c|c|c|c|c|c|c|c|c|c|} 
\hline
\multirow{2}{*}{\textbf{Classification}} & \multirow{2}{*}{\textbf{System}} & \multirow{2}{*}{\textbf{Dataset}} & \multirow{2}{*}{\textbf{Parameters}} & \multicolumn{6}{c|}{\textbf{Results}} \\ 
\cline{5-10}
 &  &  &  & \textbf{AUC} & \textbf{AUPRC} & \textbf{ACC} & \textbf{BAC} & \textbf{SEN} & \textbf{SPE} \\ 
\hline
\multirow{15}{*}{\textbf{Channel}} & \multirow{5}{*}{\textbf{TDS}} & \textbf{TUH} & \multirow{5}{*}{CC: 
    Threshold PRI} & 0.830 & 0.415 & 0.708 & 0.744 & 0.785 & 0.702 \\ 
\cline{3-3}\cline{5-10}
 &  & \textbf{NNI} &  & 0.819 & 0.698 & 0.748 & 0.733 & 0.686 & 0.779 \\ 
\cline{3-3}\cline{5-10}
 &  & \textbf{Fortis} &  & 0.633 & 0.250 & 0.652 & 0.585 & 0.477 & 0.693 \\ 
\cline{3-3}\cline{5-10}
 &  & \textbf{NUH} &  & 0.749 & 0.677 & 0.665 & 0.676 & 0.779 & 0.574 \\ 
\cline{3-3}\cline{5-10}
 &  & \textbf{Mean} &  & 0.758 & 0.510 & 0.693 & 0.684 & 0.682 & 0.687 \\ 
\cline{2-10}
 & \multirow{5}{*}{\textbf{SLDS}} & \textbf{TUH} & \multirow{5}{*}{CC: 
    LR} & 0.862 & 0.575 & 0.752 & 0.772 & 0.796 & 0.747 \\ 
\cline{3-3}\cline{5-10}
 &  & \textbf{NNI} &  & 0.857 & 0.773 & 0.782 & 0.762 & 0.694 & 0.831 \\ 
\cline{3-3}\cline{5-10}
 &  & \textbf{Fortis} &  & 0.689 & 0.309 & 0.677 & 0.632 & 0.556 & 0.708 \\ 
\cline{3-3}\cline{5-10}
 &  & \textbf{NUH} &  & 0.786 & 0.782 & 0.712 & 0.709 & 0.821 & 0.597 \\ 
\cline{3-3}\cline{5-10}
 &  & \textbf{Mean} &  & 0.798 & 0.610 & 0.731 & 0.719 & 0.717 & 0.721 \\ 
\cline{2-10}
 & \multirow{5}{*}{\textbf{DLDS}} & \textbf{TUH} & \multirow{5}{*}{CC: CNN (F:64, K:13)} & 0.827 & 0.349 & 0.655 & 0.723 & 0.806 & 0.640 \\ 
\cline{3-3}\cline{5-10}
 &  & \textbf{NNI} &  & 0.847 & 0.732 & 0.768 & 0.768 & 0.769 & 0.767 \\ 
\cline{3-3}\cline{5-10}
 &  & \textbf{Fortis} &  & 0.743 & 0.395 & 0.663 & 0.668 & 0.677 & 0.660 \\ 
\cline{3-3}\cline{5-10}
 &  & \textbf{NUH} &  & 0.791 & 0.762 & 0.720 & 0.717 & 0.845 & 0.588 \\ 
\cline{3-3}\cline{5-10}
 &  & \textbf{Mean} &  & 0.802 & 0.560 & 0.701 & 0.719 & 0.774 & 0.664 \\ 

\hline \hline

\multirow{15}{*}{\textbf{Segment}} & \multirow{5}{*}{\textbf{TDS}} & \textbf{TUH} & \multirow{5}{*}{\begin{tabular}[c]{@{}c@{}}Feature: PRI\\SC: LR\\Bins: 5\end{tabular}} & 0.761 & 0.517 & 0.732 & 0.678 & 0.581 & 0.775 \\ 
\cline{3-3}\cline{5-10}
 &  & \textbf{NNI} &  & 0.884 & 0.852 & 0.818 & 0.807 & 0.755 & 0.859 \\ 
\cline{3-3}\cline{5-10}
 &  & \textbf{Fortis} &  & 0.649 & 0.376 & 0.654 & 0.590 & 0.446 & 0.734 \\ 
\cline{3-3}\cline{5-10}
 &  & \textbf{NUH} &  & 0.758 & 0.818 & 0.691 & 0.677 & 0.782 & 0.573 \\ 
\cline{3-3}\cline{5-10}
 &  & \textbf{Mean} &  & 0.763 & 0.641 & 0.724 & 0.688 & 0.641 & 0.735 \\ 
\cline{2-10}
 & \multirow{5}{*}{\textbf{SLDS}} & \textbf{TUH} & \multirow{5}{*}{\begin{tabular}[c]{@{}c@{}}CC: LR\\SC: LR\\Bins: 2\end{tabular}} & 0.812 & 0.598 & 0.784 & 0.753 & 0.698 & 0.808 \\ 
\cline{3-3}\cline{5-10}
 &  & \textbf{NNI} &  & 0.896 & 0.868 & 0.831 & 0.821 & 0.777 & 0.866 \\ 
\cline{3-3}\cline{5-10}
 &  & \textbf{Fortis} &  & 0.694 & 0.428 & 0.692 & 0.664 & 0.6 & 0.728 \\ 
\cline{3-3}\cline{5-10}
 &  & \textbf{NUH} &  & 0.77 & 0.81 & 0.699 & 0.69 & 0.759 & 0.621 \\ 
\cline{3-3}\cline{5-10}
 &  & \textbf{Mean} &  & 0.793 & 0.676 & 0.751 & 0.732 & 0.708 & 0.756 \\ 
\cline{2-10}
 & \multirow{5}{*}{\textbf{DLDS}} & \textbf{TUH} & \multirow{5}{*}{\begin{tabular}[c]{@{}c@{}}CC: CNN (F:64, K:13)\\SC: LR\\Bins: 10\end{tabular}} & 0.767 & 0.466 & 0.745 & 0.758 & 0.783 & 0.734 \\ 
\cline{3-3}\cline{5-10}
 &  & \textbf{NNI} &  & 0.842 & 0.771 & 0.817 & 0.811 & 0.781 & 0.84 \\ 
\cline{3-3}\cline{5-10}
 &  & \textbf{Fortis} &  & 0.765 & 0.547 & 0.754 & 0.742 & 0.717 & 0.766 \\ 
\cline{3-3}\cline{5-10}
 &  & \textbf{NUH} &  & 0.783 & 0.785 & 0.725 & 0.708 & 0.815 & 0.602 \\ 
\cline{3-3}\cline{5-10}
 &  & \textbf{Mean} &  & 0.789 & 0.642 & 0.76 & 0.755 & 0.774 & 0.736 \\ 

\hline \hline

\multirow{18}{*}{\textbf{EEG}} & \multirow{6}{*}{\textbf{TDS}} & \textbf{TUH} & \multirow{6}{*}{\begin{tabular}[c]{@{}c@{}}Feature: PRI\\SC: GB\\Bins: 20\end{tabular}} & 0.95 & 0.926 & 0.923 & 0.897 & 0.96 & 0.833 \\ 
\cline{3-3}\cline{5-10}
 &  & \textbf{NNI} &  & 0.71 & 0.786 & 0.728 & 0.724 & 0.948 & 0.5 \\ 
\cline{3-3}\cline{5-10}
 &  & \textbf{Fortis} &  & 0.847 & 0.677 & 0.863 & 0.796 & 0.909 & 0.682 \\ 
\cline{3-3}\cline{5-10}
 &  & \textbf{LTMGH} &  & 0.714 & 0.637 & 0.698 & 0.611 & 0.963 & 0.26 \\ 
\cline{3-3}\cline{5-10}
 &  & \textbf{Mean} &  & 0.805 & 0.757 & 0.803 & 0.757 & 0.945 & 0.569 \\ 
\cline{3-3}\cline{5-10}
 &  & \textbf{Mean*} &  & 0.836 & 0.796 & 0.838 & 0.806 & 0.939 & 0.672 \\ 
\cline{2-10}
 & \multirow{6}{*}{\textbf{SLDS}} & \textbf{TUH} & \multirow{6}{*}{\begin{tabular}[c]{@{}c@{}}CC: SVM\\SC: LR\\Bins: 2\end{tabular}} & 0.946 & 0.895 & 0.923 & 0.911 & 0.881 & 0.941 \\ 
\cline{3-3}\cline{5-10}
 &  & \textbf{NNI} &  & 0.754 & 0.763 & 0.702 & 0.700 & 0.607 & 0.793 \\ 
\cline{3-3}\cline{5-10}
 &  & \textbf{Fortis} &  & 0.838 & 0.664 & 0.790 & 0.781 & 0.765 & 0.797 \\ 
\cline{3-3}\cline{5-10}
 &  & \textbf{LTMGH} &  & 0.713 & 0.570 & 0.423 & 0.539 & 0.964 & 0.114 \\ 
\cline{3-3}\cline{5-10}
 &  & \textbf{Mean} &  & 0.813 & 0.723 & 0.710 & 0.733 & 0.804 & 0.661 \\ 
\cline{3-3}\cline{5-10}
 &  & \textbf{Mean*} &  & 0.846 & 0.774 & 0.805 & 0.797 & 0.751 & 0.844 \\ 
\cline{2-10}
 & \multirow{6}{*}{\textbf{DLDS}} & \textbf{TUH} & \multirow{6}{*}{\begin{tabular}[c]{@{}c@{}}CC: CNN (F:64, K:9)\\SC: LR\\Bins: 10\end{tabular}} & 0.961 & 0.919 & 0.901 & 0.916 & 0.952 & 0.879 \\ 
\cline{3-3}\cline{5-10}
 &  & \textbf{NNI} &  & 0.728 & 0.778 & 0.728 & 0.726 & 0.589 & 0.862 \\ 
\cline{3-3}\cline{5-10}
 &  & \textbf{Fortis} &  & 0.847 & 0.674 & 0.855 & 0.817 & 0.753 & 0.882 \\ 
\cline{3-3}\cline{5-10}
 &  & \textbf{LTMGH} &  & 0.598 & 0.390 & 0.376 & 0.506 & 0.982 & 0.030 \\ 
\cline{3-3}\cline{5-10}
 &  & \textbf{Mean} &  & 0.783 & 0.690 & 0.715 & 0.741 & 0.819 & 0.663 \\ 
\cline{3-3}\cline{5-10}
 &  & \textbf{Mean*} &  & 0.845 & 0.790 & 0.828 & 0.820 & 0.765 & 0.874 \\
\hline
\end{tabular}

\begin{tablenotes}
\setlength\labelsep{0pt}
\footnotesize
\item CC: Channel classifier, SC: Segment/EEG classifier, Bins: Histogram bins.
\item ACC: Accuracy, BAC: Balanced Accuracy, SEN: Sensitivity, SPE: Specificity, F: Number of filters, K: Filter size.
\item *: Excluding the LTMGH dataset.
\end{tablenotes}

\end{threeparttable}
}}
\vspace{2cm}
\end{table*}

\begin{table*}[tp!]
\centering
\tbl{Channel, segment, and EEG-level LOSO CV results for the different datasets.\label{tab:LOSO}}
{
\scalebox{1.0}{
\begin{threeparttable}

\begin{tabular}{|c|c|c|c|c|c|c|c|c|c|} 
\hline
\multirow{2}{*}{\textbf{Classification}} & \multirow{2}{*}{\textbf{System}} & \multirow{2}{*}{\textbf{Dataset}} & \multirow{2}{*}{\textbf{Parameters}} & \multicolumn{6}{c|}{\textbf{Results}} \\ 
\cline{5-10}
 &  &  &  & \textbf{AUC} & \textbf{AUPRC} & \textbf{ACC} & \textbf{BAC} & \textbf{SEN} & \textbf{SPE} \\ 
\hline
\multirow{15}{*}{\textbf{Channel}} & \multirow{5}{*}{\textbf{TDS}} & \textbf{TUH} & \multirow{5}{*}{CC:
  Threshold PRI} & 0.830 & 0.415 & 0.837 & 0.763 & 0.676 & 0.850 \\ 
\cline{3-3}\cline{5-10}
 &  & \textbf{NNI} &  & 0.819 & 0.698 & 0.736 & 0.738 & 0.745 & 0.731 \\ 
\cline{3-3}\cline{5-10}
 &  & \textbf{Fortis} &  & 0.633 & 0.250 & 0.545 & 0.614 & 0.726 & 0.502 \\ 
\cline{3-3}\cline{5-10}
 &  & \textbf{NUH} &  & 0.749 & 0.677 & 0.690 & 0.685 & 0.648 & 0.723 \\ 
\cline{3-3}\cline{5-10}
 &  & \textbf{Mean} &  & 0.758 & 0.510 & 0.702 & 0.700 & 0.699 & 0.702 \\ 
\cline{2-10}
 & \multirow{5}{*}{\textbf{SLDS}} & \textbf{TUH} & \multirow{5}{*}{CC:
  SVM\_rbf} & 0.676 & 0.174 & 0.826 & 0.677 & 0.496 & 0.858 \\ 
\cline{3-3}\cline{5-10}
 &  & \textbf{NNI} &  & 0.828 & 0.695 & 0.773 & 0.776 & 0.788 & 0.764 \\ 
\cline{3-3}\cline{5-10}
 &  & \textbf{Fortis} &  & 0.626 & 0.308 & 0.648 & 0.602 & 0.525 & 0.679 \\ 
\cline{3-3}\cline{5-10}
 &  & \textbf{NUH} &  & 0.759 & 0.737 & 0.707 & 0.707 & 0.677 & 0.738 \\ 
\cline{3-3}\cline{5-10}
 &  & \textbf{Mean} &  & 0.722 & 0.479 & 0.739 & 0.691 & 0.622 & 0.760 \\ 
\cline{2-10}
 & \multirow{5}{*}{\textbf{DLDS}} & \textbf{TUH} & \multirow{5}{*}{CC:
  CNN (F:32, K:7)} & 0.791 & 0.237 & 0.715 & 0.762 & 0.820 & 0.704 \\ 
\cline{3-3}\cline{5-10}
 &  & \textbf{NNI} &  & 0.837 & 0.667 & 0.738 & 0.765 & 0.856 & 0.674 \\ 
\cline{3-3}\cline{5-10}
 &  & \textbf{Fortis} &  & 0.725 & 0.390 & 0.621 & 0.655 & 0.713 & 0.598 \\ 
\cline{3-3}\cline{5-10}
 &  & \textbf{NUH} &  & 0.804 & 0.812 & 0.718 & 0.715 & 0.824 & 0.606 \\ 
\cline{3-3}\cline{5-10}
 &  & \textbf{Mean} &  & 0.789 & 0.527 & 0.698 & 0.724 & 0.803 & 0.646 \\ 
 
\hline \hline

\multirow{15}{*}{\textbf{Segment}} & \multirow{5}{*}{\textbf{TDS}} & \textbf{TUH} & \multirow{5}{*}{\begin{tabular}[c]{@{}c@{}}Feature: PRI\\SC: LR\\Bins: 5\end{tabular}} & 0.827 & 0.690 & 0.809 & 0.769 & 0.698 & 0.841 \\ 
\cline{3-3}\cline{5-10}
 &  & \textbf{NNI} &  & 0.858 & 0.818 & 0.775 & 0.758 & 0.670 & 0.845 \\ 
\cline{3-3}\cline{5-10}
 &  & \textbf{Fortis} &  & 0.689 & 0.539 & 0.692 & 0.669 & 0.615 & 0.722 \\ 
\cline{3-3}\cline{5-10}
 &  & \textbf{NUH} &  & 0.692 & 0.749 & 0.644 & 0.658 & 0.549 & 0.767 \\ 
\cline{3-3}\cline{5-10}
 &  & \textbf{Mean} &  & 0.766 & 0.699 & 0.730 & 0.713 & 0.633 & 0.794 \\ 
\cline{2-10}
 & \multirow{5}{*}{\textbf{SLDS}} & \textbf{TUH} & \multirow{5}{*}{\begin{tabular}[c]{@{}c@{}}CC: SVM\_rbf\\SC: RF\\Bins: 5\end{tabular}} & 0.745 & 0.491 & 0.742 & 0.710 & 0.651 & 0.768 \\ 
\cline{3-3}\cline{5-10}
 &  & \textbf{NNI} &  & 0.845 & 0.732 & 0.822 & 0.818 & 0.798 & 0.838 \\ 
\cline{3-3}\cline{5-10}
 &  & \textbf{Fortis} &  & 0.586 & 0.351 & 0.650 & 0.582 & 0.431 & 0.734 \\ 
\cline{3-3}\cline{5-10}
 &  & \textbf{NUH} &  & 0.703 & 0.760 & 0.661 & 0.671 & 0.594 & 0.748 \\ 
\cline{3-3}\cline{5-10}
 &  & \textbf{Mean} &  & 0.720 & 0.584 & 0.719 & 0.695 & 0.619 & 0.772 \\ 
\cline{2-10}
 & \multirow{5}{*}{\textbf{DLDS}} & \textbf{TUH} & \multirow{5}{*}{\begin{tabular}[c]{@{}c@{}}CC: CNN (F:32, K:7)\\SC: LR\\Bins: 2\end{tabular}} & 0.749 & 0.511 & 0.825 & 0.780 & 0.696 & 0.864 \\ 
\cline{3-3}\cline{5-10}
 &  & \textbf{NNI} &  & 0.851 & 0.772 & 0.829 & 0.832 & 0.844 & 0.819 \\ 
\cline{3-3}\cline{5-10}
 &  & \textbf{Fortis} &  & 0.747 & 0.455 & 0.723 & 0.715 & 0.698 & 0.731 \\ 
\cline{3-3}\cline{5-10}
 &  & \textbf{NUH} &  & 0.748 & 0.745 & 0.742 & 0.737 & 0.770 & 0.704 \\ 
\cline{3-3}\cline{5-10}
 &  & \textbf{Mean} &  & 0.774 & 0.621 & 0.780 & 0.766 & 0.752 & 0.780 \\ 
 
\hline \hline

\multirow{18}{*}{\textbf{EEG}} & \multirow{6}{*}{\textbf{TDS}} & \textbf{TUH} & \multirow{6}{*}{\begin{tabular}[c]{@{}c@{}}Feature: 4 RP\\SC: GB\\Bins: 20\end{tabular}} & 0.942 & 0.906 & 0.923 & 0.911 & 0.941 & 0.881 \\ 
\cline{3-3}\cline{5-10}
 &  & \textbf{NNI} &  & 0.76 & 0.775 & 0.746 & 0.744 & 0.828 & 0.661 \\ 
\cline{3-3}\cline{5-10}
 &  & \textbf{Fortis} &  & 0.846 & 0.706 & 0.872 & 0.806 & 0.918 & 0.694 \\ 
\cline{3-3}\cline{5-10}
 &  & \textbf{LTMGH} &  & 0.829 & 0.72 & 0.762 & 0.758 & 0.775 & 0.74 \\ 
\cline{3-3}\cline{5-10}
 &  & \textbf{Mean} &  & 0.844 & 0.777 & 0.826 & 0.805 & 0.866 & 0.744 \\ 
\cline{3-3}\cline{5-10}
 &  & \textbf{Mean*} &  & 0.849 & 0.796 & 0.847 & 0.820 & 0.896 & 0.745 \\ 
\cline{2-10}
 & \multirow{6}{*}{\textbf{SLDS}} & \textbf{TUH} & \multirow{6}{*}{\begin{tabular}[c]{@{}c@{}}CC: RF\\SC: LR\\Bins: 5\end{tabular}} & 0.919 & 0.897 & 0.895 & 0.884 & 0.857 & 0.911 \\ 
\cline{3-3}\cline{5-10}
 &  & \textbf{NNI} &  & 0.828 & 0.844 & 0.772 & 0.771 & 0.714 & 0.828 \\ 
\cline{3-3}\cline{5-10}
 &  & \textbf{Fortis} &  & 0.831 & 0.641 & 0.863 & 0.804 & 0.706 & 0.903 \\ 
\cline{3-3}\cline{5-10}
 &  & \textbf{LTMGH} &  & 0.743 & 0.609 & 0.732 & 0.716 & 0.657 & 0.774 \\ 
\cline{3-3}\cline{5-10}
 &  & \textbf{Mean} &  & 0.830 & 0.748 & 0.815 & 0.794 & 0.734 & 0.854 \\ 
\cline{3-3}\cline{5-10}
 &  & \textbf{Mean*} &  & 0.859 & 0.794 & 0.843 & 0.820 & 0.759 & 0.881 \\ 
\cline{2-10}
 & \multirow{6}{*}{\textbf{DLDS}} & \textbf{TUH} & \multirow{6}{*}{\begin{tabular}[c]{@{}c@{}}CC: CNN (F:32, K:5)\\SC: LR\\Bins: 15\end{tabular}} & 0.943 & 0.853 & 0.922 & 0.917 & 0.905 & 0.929 \\ 
\cline{3-3}\cline{5-10}
 &  & \textbf{NNI} &  & 0.774 & 0.801 & 0.754 & 0.751 & 0.571 & 0.931 \\ 
\cline{3-3}\cline{5-10}
 &  & \textbf{Fortis} &  & 0.836 & 0.652 & 0.841 & 0.786 & 0.694 & 0.879 \\ 
\cline{3-3}\cline{5-10}
 &  & \textbf{LTMGH} &  & 0.723 & 0.573 & 0.704 & 0.690 & 0.639 & 0.741 \\ 
\cline{3-3}\cline{5-10}
 &  & \textbf{Mean} &  & 0.819 & 0.720 & 0.805 & 0.786 & 0.702 & 0.870 \\ 
\cline{3-3}\cline{5-10}
 &  & \textbf{Mean*} &  & 0.851 & 0.769 & 0.839 & 0.818 & 0.723 & 0.913 \\
\hline
\end{tabular}

\begin{tablenotes}
\setlength\labelsep{0pt}
\footnotesize
\item CC: Channel classifier, SC: Segment/EEG classifier, Bins: Histogram bins.
\item ACC: Accuracy, BAC: Balanced Accuracy, SEN: Sensitivity, SPE: Specificity, F: Number of filters, K: Filter size.
\item *: Excluding the LTMGH dataset.
\end{tablenotes}

\end{threeparttable}
}}
\vspace{2cm}
\end{table*}

\subsection{Threshold-based EEG-level classification}

\begin{figure*}[tb]
\begin{minipage}[b]{0.32\linewidth}
  \centering
  \centerline{\includegraphics[width=6cm]{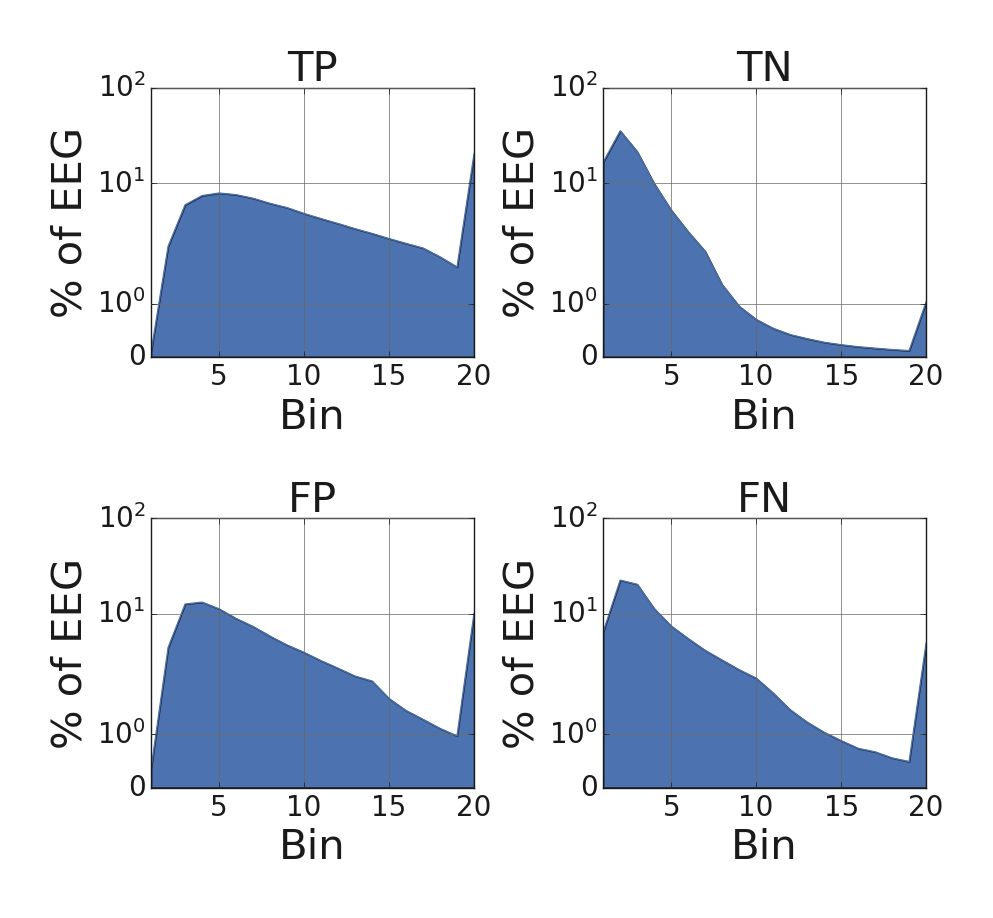}}
  \vspace{-0.3 cm}
  \centerline{{\tabular[t]{@{}l@{}}(a) PRI values for TDS. \\ \phantom{.} \\ \endtabular}}\medskip
\end{minipage}
\hfill
\begin{minipage}[b]{0.32\linewidth}
  \centering
  \centerline{\includegraphics[width=6cm]{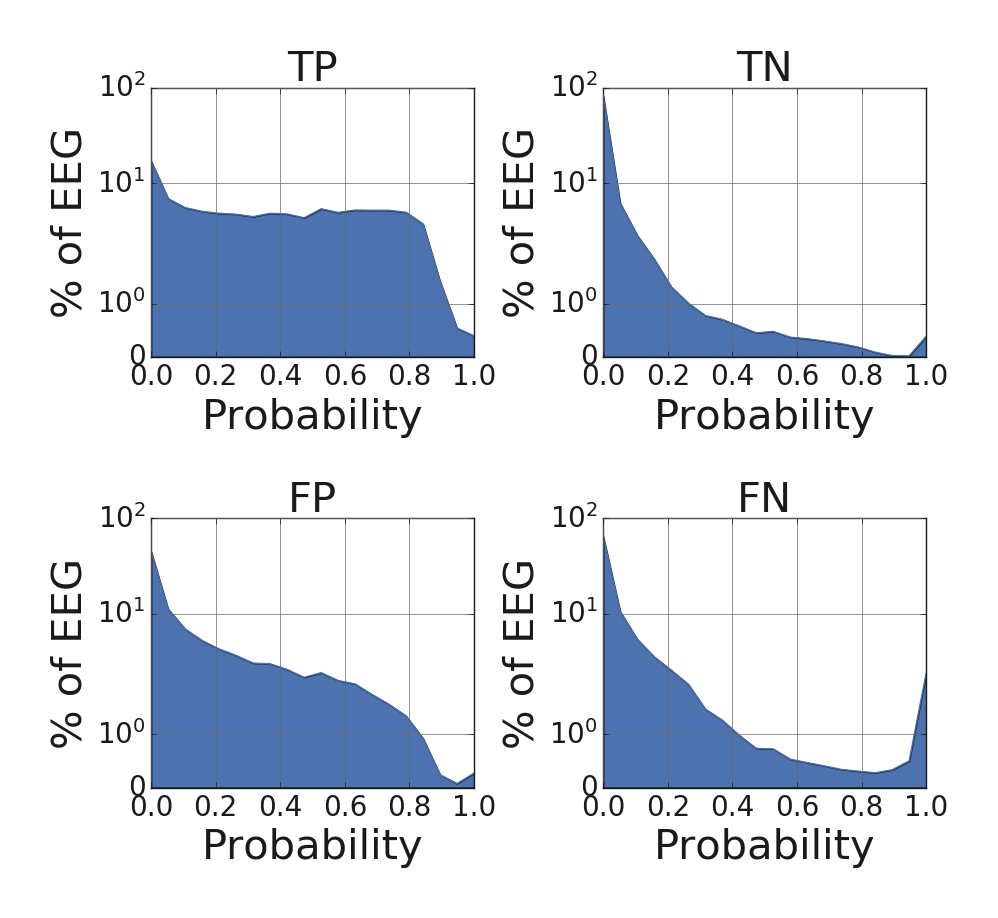}}
  \vspace{-0.3 cm}
  \centerline{{\tabular[t]{@{}l@{}}(b) Channel-level slowing \\ detector outputs for SLDS. \endtabular}}\medskip
\end{minipage}
\hfill
\begin{minipage}[b]{0.32\linewidth}
  \centering
  \centerline{\includegraphics[width=6cm]{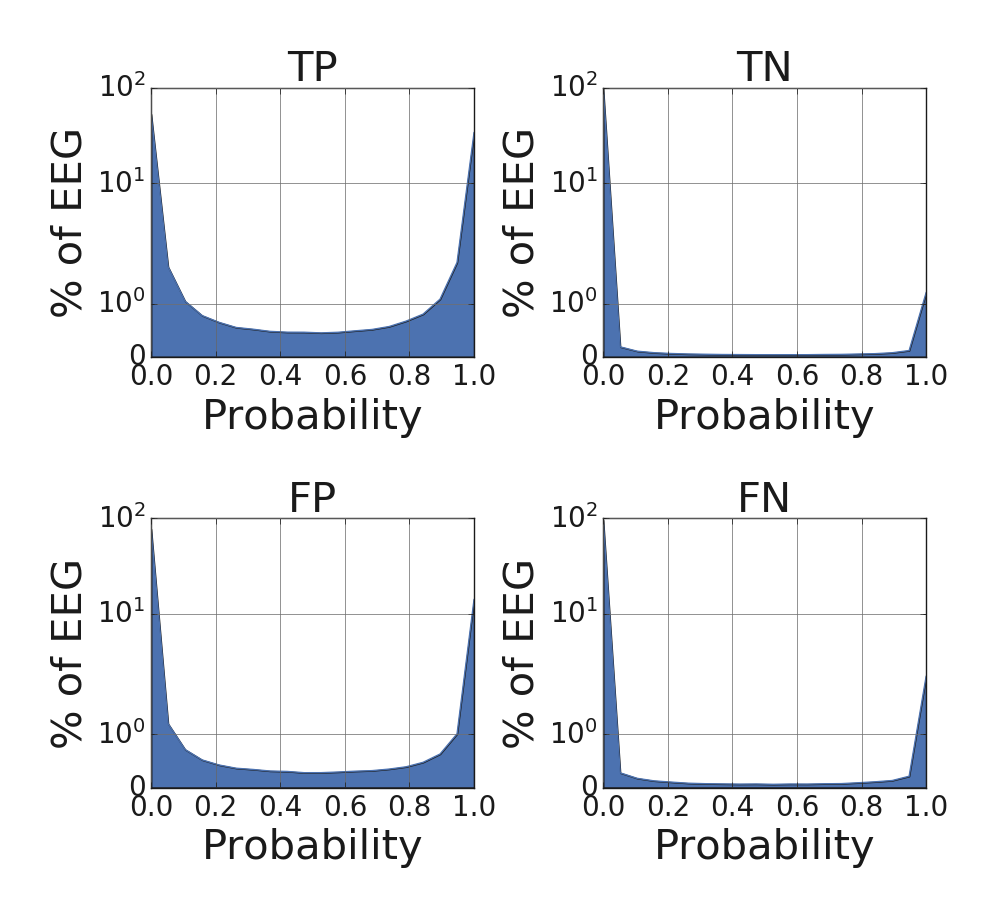}}
  \vspace{-0.3 cm}
  \centerline{{\tabular[t]{@{}l@{}}(c) Channel-level slowing \\ detector outputs for DLDS. \endtabular}}\medskip
\end{minipage}
\vspace{-0.3 cm}
\caption{Distribution of PRI values for TDS (a) and of channel-level slowing detector outputs for SLDS (b) and DLDS (c). Distributions for the TP, FN, FP, and FN of the classification results are displayed. The y-axis is in symmetric log scale.}
\label{fig:histogram_distribution}
\vspace{-0.4 cm}
\end{figure*}

\begin{figure*}[tb]
\begin{minipage}[b]{0.32\linewidth}
  \centering
  \centerline{\includegraphics[width=6cm]{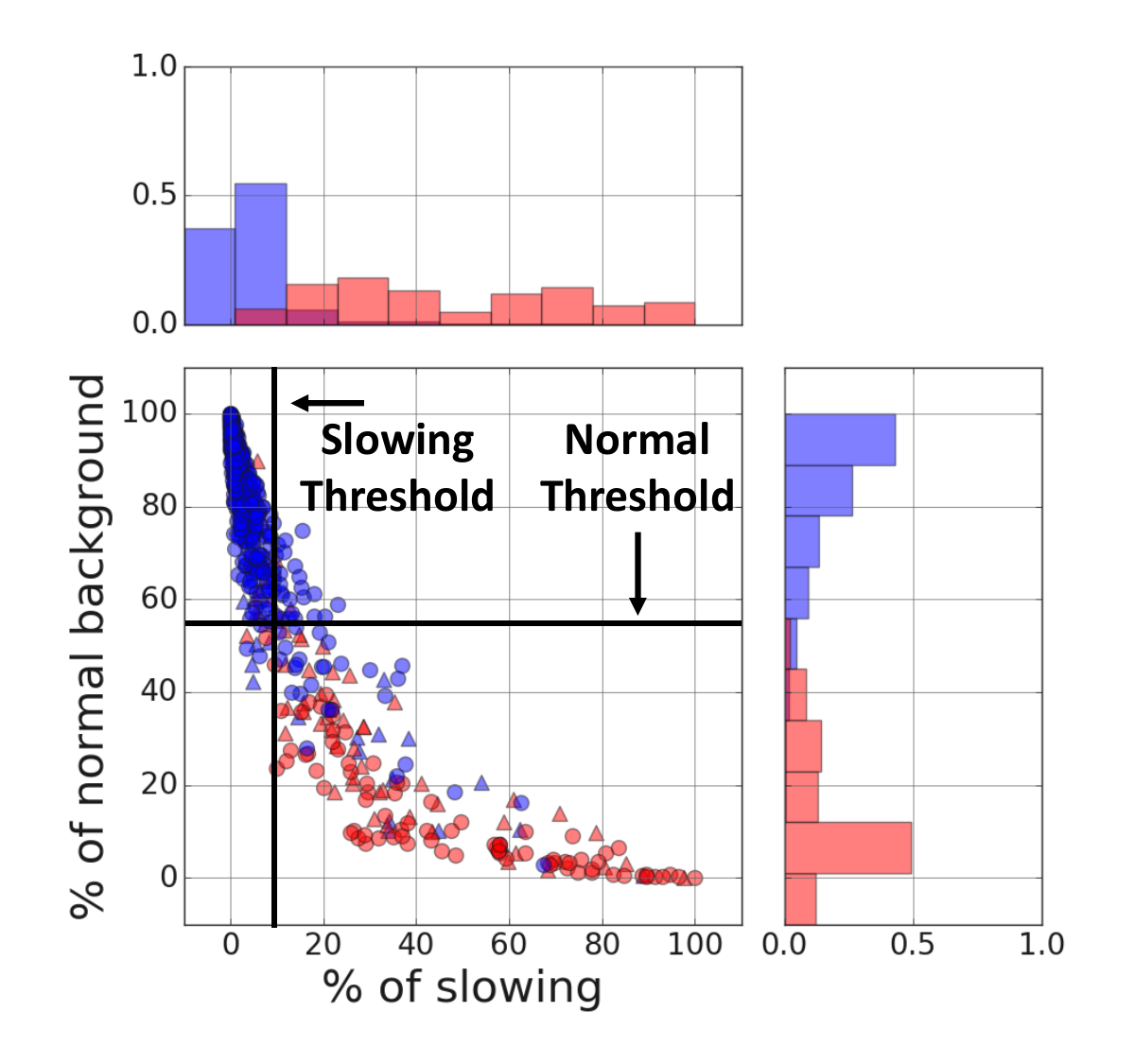}}
  \vspace{-0.3 cm}
  \centerline{(a) TDS scatterplot.}\medskip
\end{minipage}
\hfill
\begin{minipage}[b]{0.32\linewidth}
  \centering
  \centerline{\includegraphics[width=6cm]{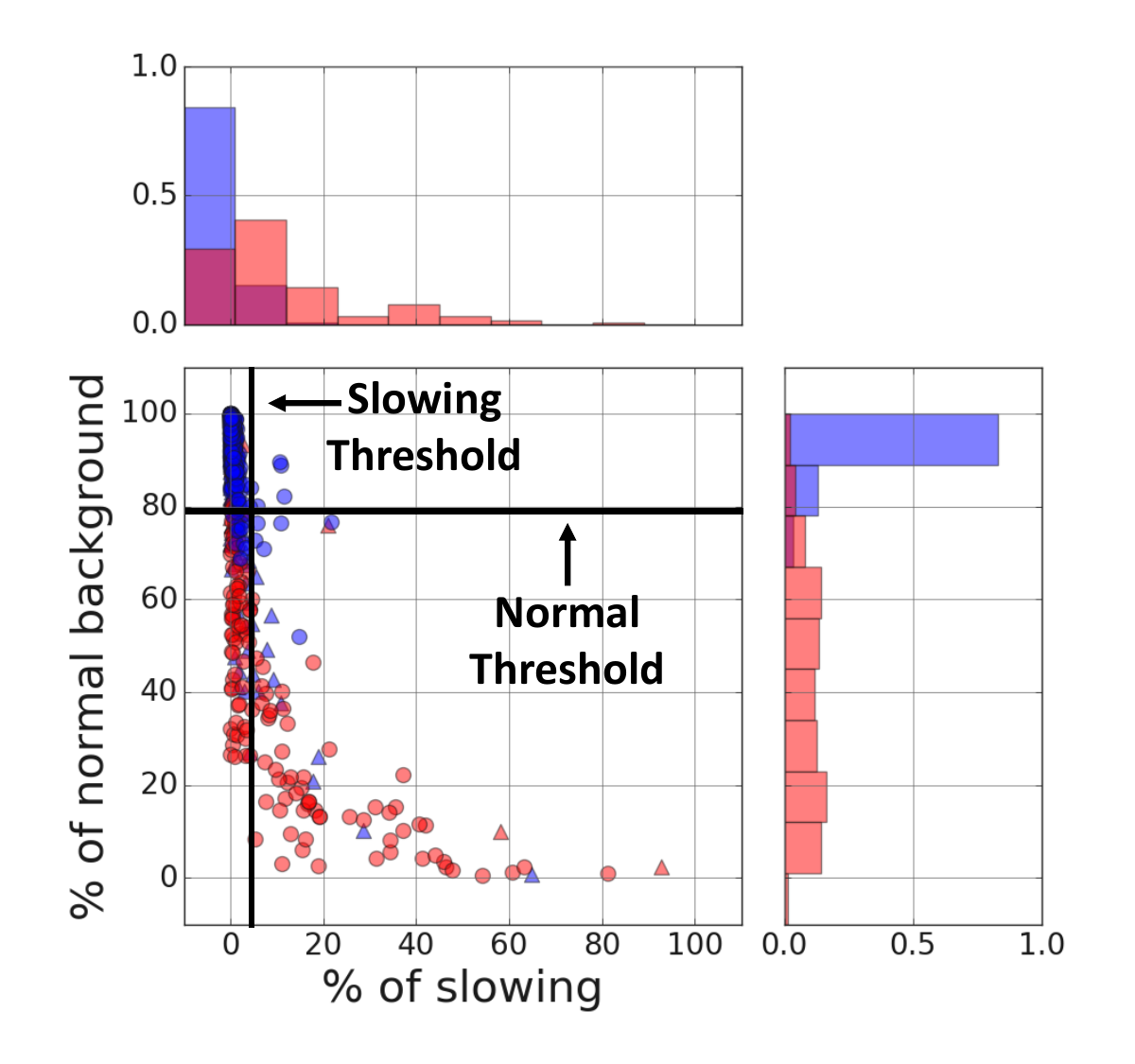}}
  \vspace{-0.3 cm}
  \centerline{(b) SLDS scatterplot.}\medskip
\end{minipage}
\hfill
\begin{minipage}[b]{0.32\linewidth}
  \centering
  \centerline{\includegraphics[width=6cm]{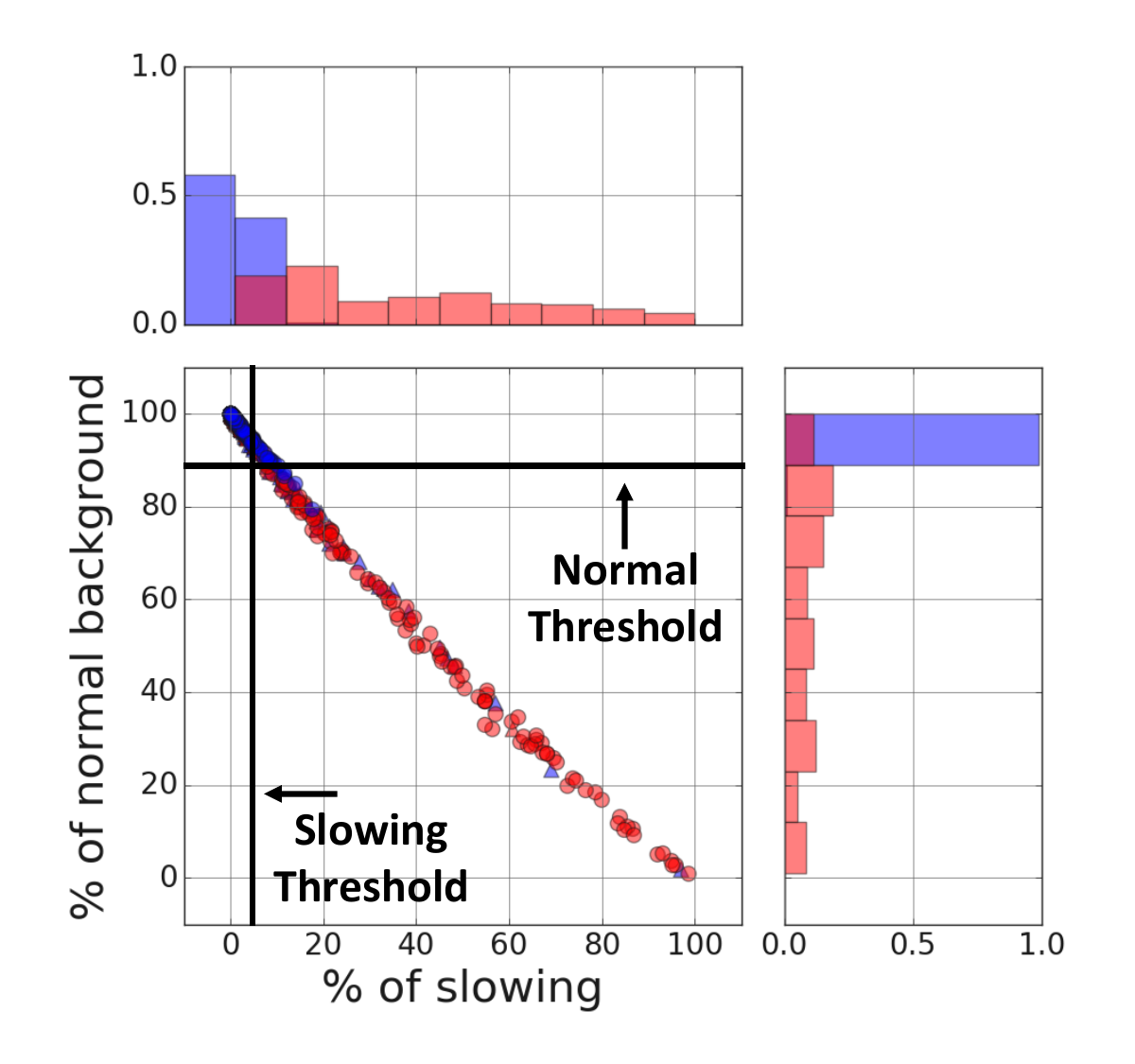}}
  \vspace{-0.3 cm}
  \centerline{(c) DLDS scatterplot.}\medskip
\end{minipage}
\vspace{-0.3 cm}
\caption{EEG-level slowing versus background percentage scatterplot. a) TDS, b) SLDS, c) DLDS. Slow-free and slowing EEGs are denoted in blue and red, respectively. The normalized histogram distribution of the percentage of normal and slowing segments is illustrated on the top and right sides of each scatterplot. From the scatterplot, we can determine the best threshold.}
\label{fig:slow_threshold}
\vspace{-0.2 cm}
\end{figure*}

In this section, we show how shallow learning models can accurately detect EEG slowing through histogram-based features. The TDS deploys spectral features (the PRI is selected for illustration, as it yields the best results), while the SLDS and DLDS rely on histograms of the outputs of the channel-level slowing detector. We plot the average histogram distribution (20 bins) of all EEGs across the datasets (TUH, NNI, Fortis) in Figure \ref{fig:histogram_distribution}. The histograms are split based on the EEG-level LOIO CV classification results of the respective systems: $TN$, $TN$, $FP$, and $FN$. 

The histograms show differences across slow-free and slowing EEGs for the three systems. The TDS detects pathological slowing in an EEG if it detects a high percentage of single-channel segments with high PRI values. On the other hand, the SLDS and DLDS detect abnormal slowing in an EEG when the output of the single-channel slowing detector is frequently close to 1. The histogram from the DLDS is more skewed than those in the SLDS. To compare the slowing and slow-free EEGs, we define a normal EEG background segment between bin 1 to 5 and a slow EEG segment between bin 15 to 20. The bins between 6 and 14 are not included. The PRI values and slowing detector outputs corresponding to the bin numbers are listed in Table \ref{tab:bin_range}. Figure \ref{fig:slow_threshold} shows scatterplots of the percentage of normal EEG background versus slowing percentage.

\begin{tablehere}
\vspace{-0.4cm}
\centering
\tbl{Histogram bins and their corresponding PRI and slowing detector outputs. \label{tab:bin_range}}
{
\scalebox{0.9}{
\begin{threeparttable}
\begin{tabular}{|c|c|c|} 
\cline{2-3}
\multicolumn{1}{c|}{} & \textbf{TDS} & \textbf{SLDS and DLDS} \\ 
\hline
\textbf{Bin range} & \textbf{PRI range} & \textbf{Slowing detector output range} \\ 
\hline
{[}1,5] & 0-2.822 & 0-0.25 \\
{[}6,14] & 2.822-9.984 & 0.25-0.75 \\
{[}15,20] & 9.984 & 0.75-1 \\
\hline
\end{tabular}
\end{threeparttable}
}}
\vspace{-0.4 cm}
\end{tablehere}

The scatterplot for the TDS and SLDS displayed a non-linear pattern, while for the DLDS, it generated a linear trend. The linearity of the DLDS scatterplot is due to the skewed output distribution of the slowing detector (see Figure \ref{fig:histogram_distribution}). The slow-free and slowing EEG exhibit clear distinctions for the three systems, enabling us to apply thresholding on the percentage of normal background or slowing duration to classify the EEGs. We tested all histogram bins as potential thresholds for the binary classification. For each system, we utilize different threshold to compute the classification BAC for each dataset (TUH, NNI, Fortis), and take the mean BAC. The thresholds associated with the highest mean BAC for each system are listed in Table \ref{tab:normal_slow_threshold_table}. 

As both the SLDS and DLDS leverage the channel-level slowing detector outputs histograms, they have similar optimal thresholds. The TDS has the best overall mean BAC of 80.9\% with a 55\% threshold on the percentage of the normal background. The SLDS and DLDS achieve a mean BAC of 80.0\% and 78.0\% with a threshold on the normal background percentage set at 80\% and 90\%, respectively. A threshold on the normal background percentage yields better classification results than a threshold on the slow percentage. Thresholding is more interpretable and can yield comparable results to the EEG-level LOIO and LOSO CV with the shallow learning model.

\begin{tablehere}
\vspace{-0.4cm}
\centering
\tbl{EEG-level classification mean BAC with threshold. \label{tab:normal_slow_threshold_table}}
{
\scalebox{0.9}{
\begin{threeparttable}
\begin{tabular}{|c|c|c|c|c|} 
\hline
\multirow{2}{*}{\textbf{System}} & \multicolumn{2}{c|}{\textbf{Normal Background \%}} & \multicolumn{2}{c|}{\textbf{Slowing \%}} \\ 
\cline{2-5}
 & \textbf{Threshold}  & \textbf{BAC}  & \textbf{Threshold}  & \textbf{BAC}  \\ 
\hline
\textbf{TDS}  & 55 & 0.809 & 10 & 0.786 \\ 
\hline
\textbf{SLDS}  & 80 & 0.800 & 5 & 0.649 \\ 
\hline
\textbf{DLDS}  & 90 & 0.780 & 5 & 0.777 \\
\hline
\end{tabular}
\end{threeparttable}
}}
\vspace{-0.4 cm}
\end{tablehere}

\begin{figure*}[tb]
\begin{minipage}[b]{0.32\linewidth}
  \centering
  \centerline{\includegraphics[width=6cm]{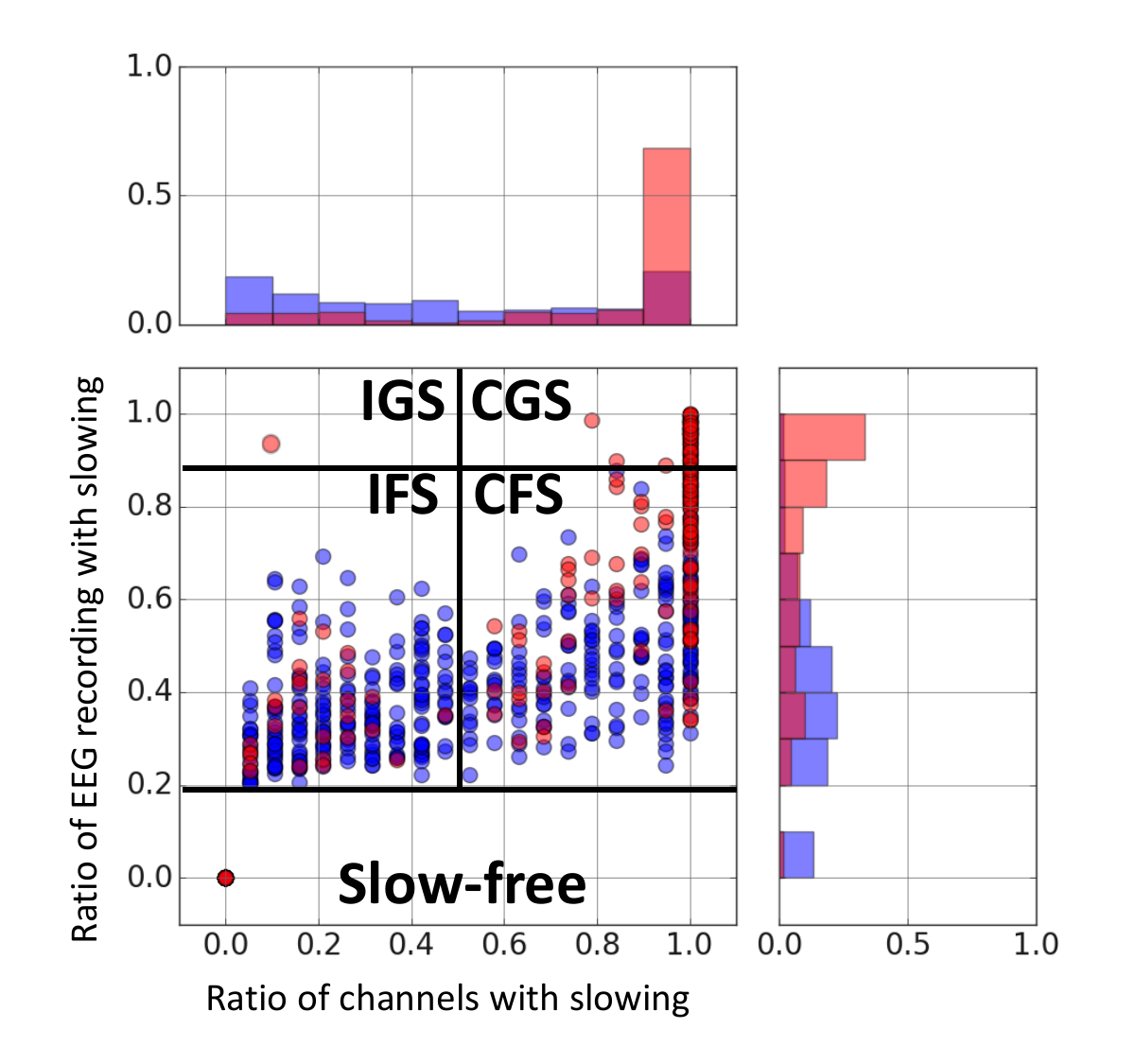}}
  \vspace{-0.3 cm}
  \centerline{(a) Slowing detected by the TDS.}\medskip
\end{minipage}
\hfill
\begin{minipage}[b]{0.32\linewidth}
  \centering
  \centerline{\includegraphics[width=6cm]{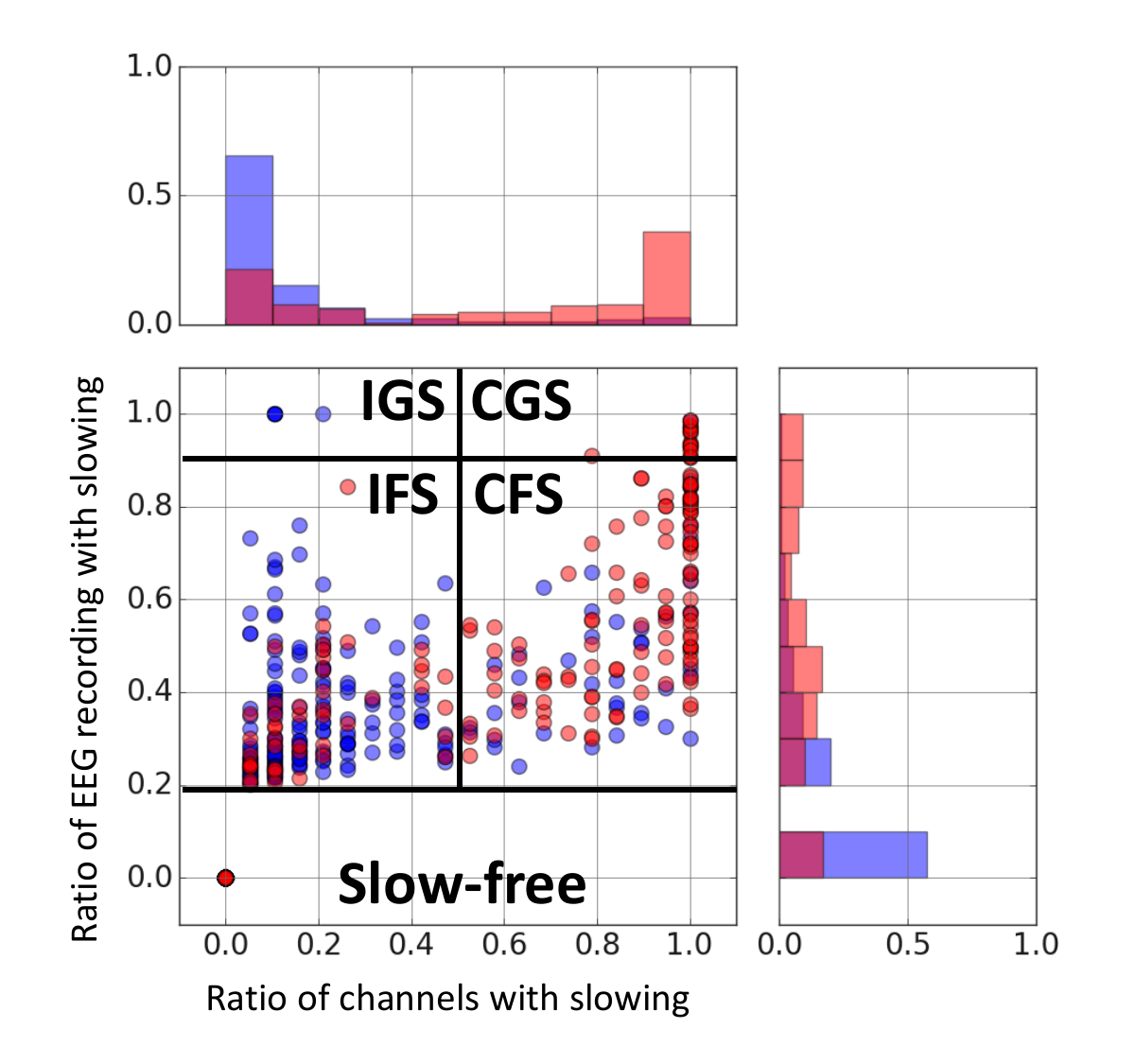}}
  \vspace{-0.3 cm}
  \centerline{(b) Slowing detected by the SLDS.}\medskip
\end{minipage}
\hfill
\begin{minipage}[b]{0.32\linewidth}
  \centering
  \centerline{\includegraphics[width=6cm]{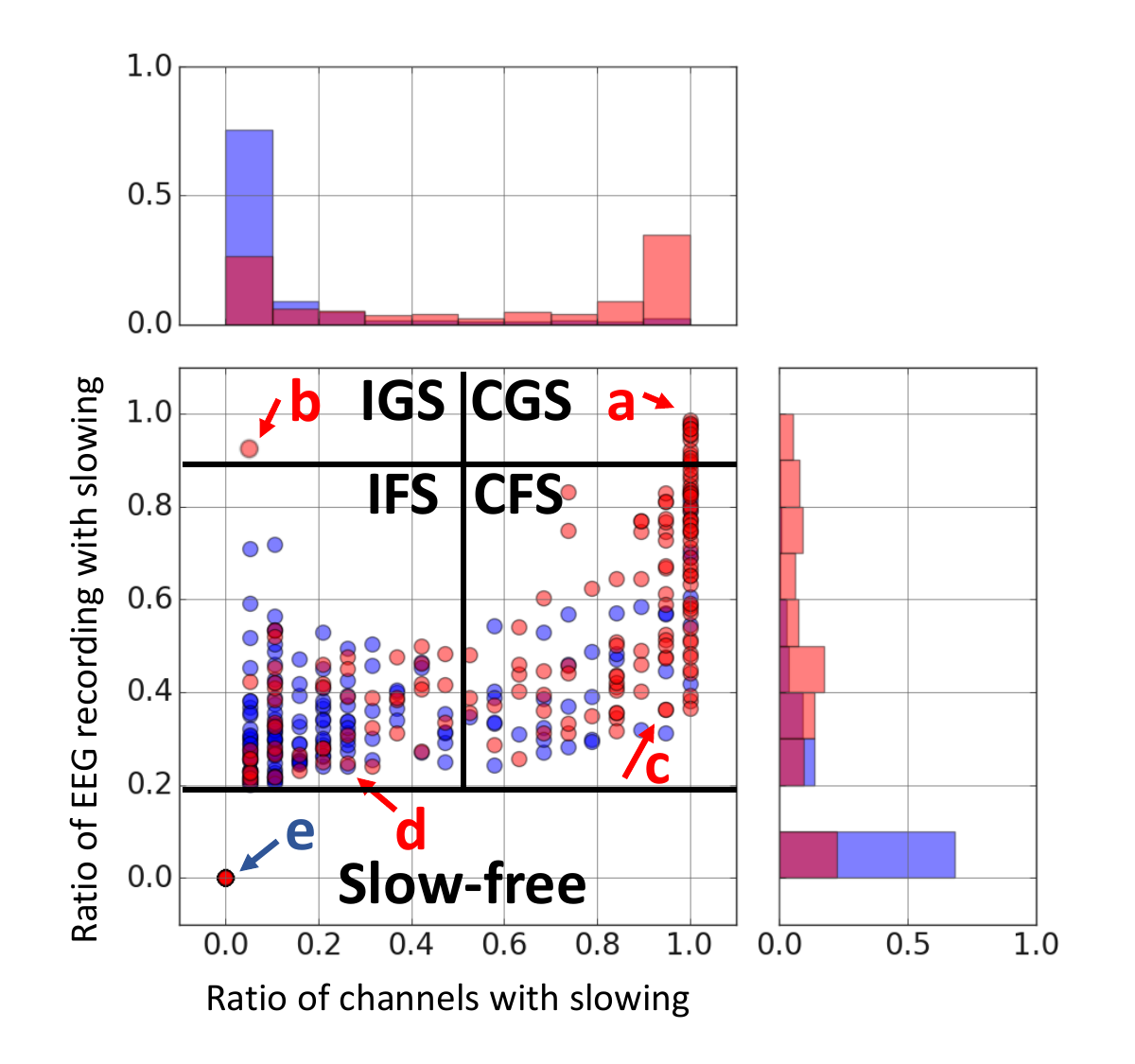}}
  \vspace{-0.3 cm}
  \centerline{(c) Slowing detected by the DLDS.}\medskip
\end{minipage}
\vspace{-0.3 cm}
\caption{Four degrees of slowing (CGS, IGS, CFS, and IFS) were detected in EEG-level for the DLDS. Each blue and red dot represents a slow-free EEG and an EEG with pathological slowing. We display an example of CGS, IGS, CFS, IFS and a slow-free EEG detected by DLDS in Figure \ref{fig:type_of_slow}.}
\label{fig:degree_of_slowing}
\vspace{-0.3 cm}
\end{figure*}

\ExplSyntaxOn
\cs_set_eq:NN \atdot_centerline:n \centerline
\cs_set_protected_nopar:Npn \centerline #1
 {
  \seq_set_split:Nnn \l_atdotde_centerline_seq { \\ } { #1 }
  \seq_map_inline:Nn \l_atdotde_centerline_seq
   {
    \atdot_centerline:n { ##1 }
   }
 }
\ExplSyntaxOff

\begin{figure*}
\hfill
\begin{minipage}[b]{0.19\linewidth}
  \centering
  \centerline{\includegraphics[width=5cm]{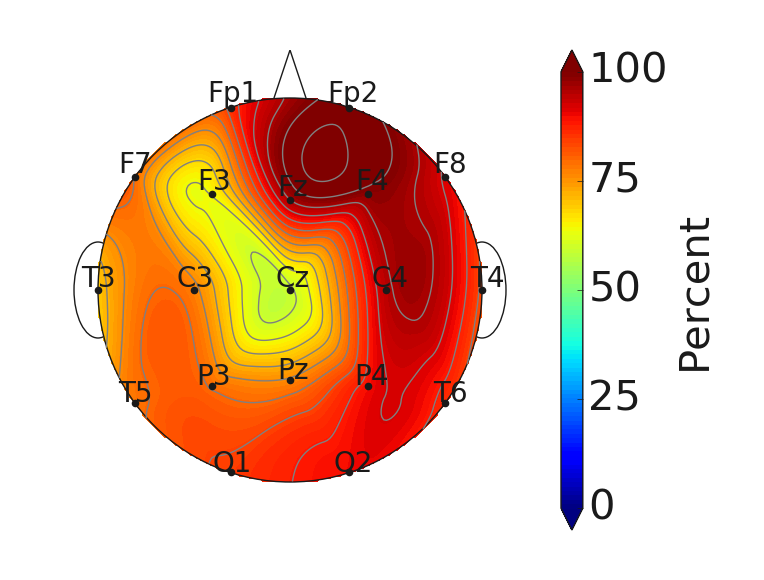}}
  \vspace{-0.1cm}
  \centerline{(a) Continuous and\\generalized slowing.}\medskip
\end{minipage}
\hfill
\begin{minipage}[b]{0.19\linewidth}
  \centering
  \centerline{\includegraphics[width=5cm]{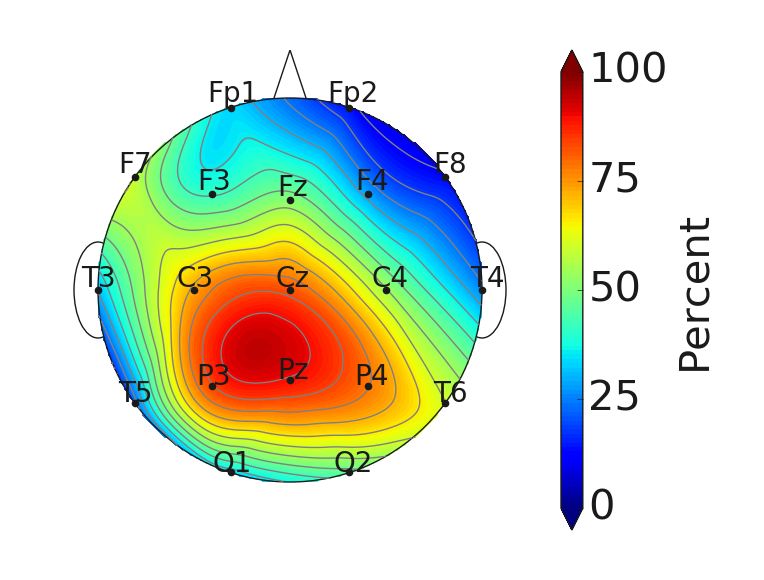}}
  \vspace{-0.1 cm}
  \centerline{(b) Intermittent and\\generalized slowing.}\medskip
\end{minipage}
\hfill
\begin{minipage}[b]{0.19\linewidth}
  \centering
  \centerline{\includegraphics[width=5cm]{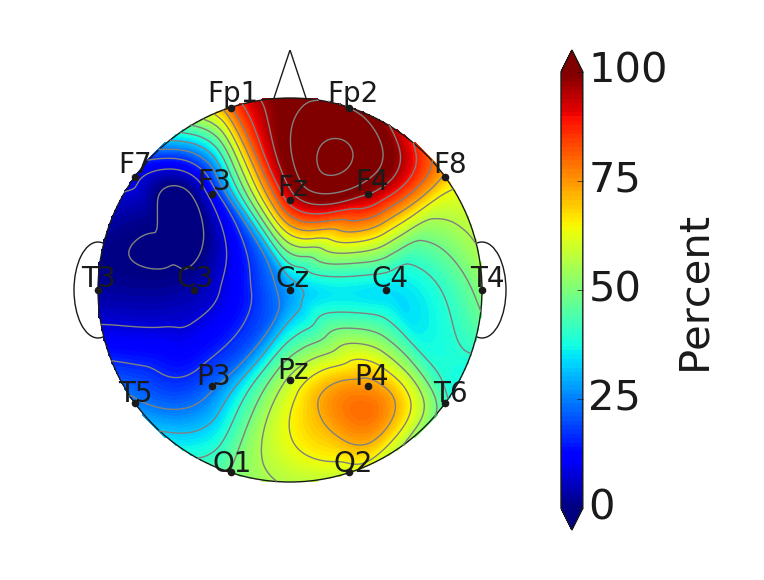}}
  \vspace{-0.1 cm}
  \centerline{(c) Continuous and\\focal slowing.}\medskip
\end{minipage}
\hfill
\begin{minipage}[b]{0.19\linewidth}
  \centering
  \centerline{\includegraphics[width=5cm]{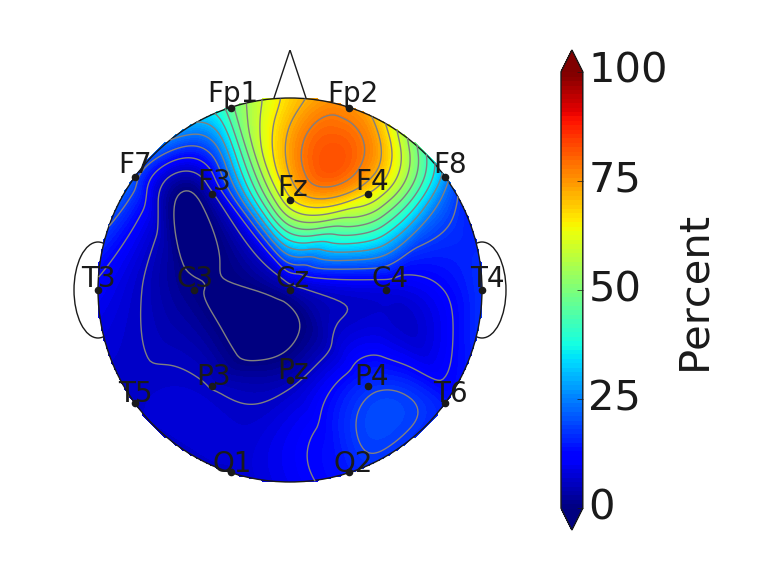}}
  \vspace{-0.1 cm}
  \centerline{(d) Intermittent and\\focal slowing.}\medskip
\end{minipage}
\hfill
\begin{minipage}[b]{0.19\linewidth}
  \centering
  \centerline{\includegraphics[width=5cm]{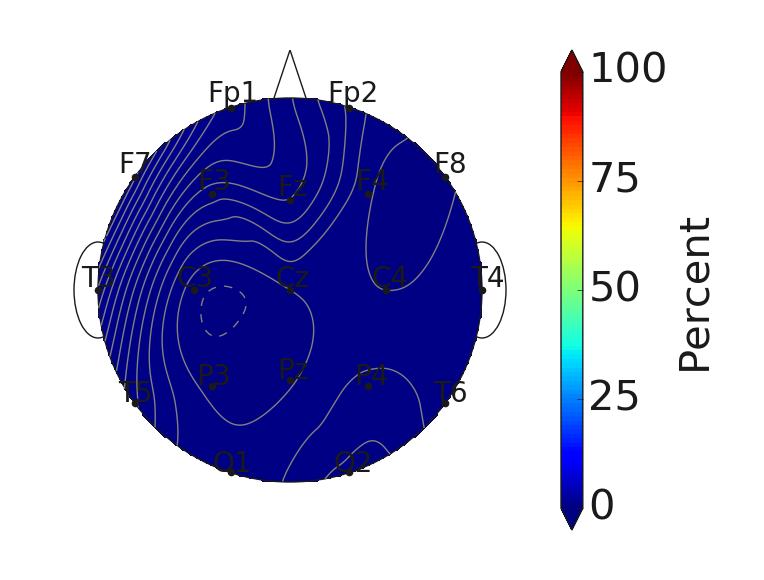}}
  \vspace{-0.1 cm}
  \centerline{(e) Slow-free EEG. \\}\medskip
\end{minipage}
\hfill
\vspace{-0.1cm}
\caption{Examples of EEGs with different degrees of slowing and a slow-free EEG. The percentage of slowing for each channel is displayed.}
\label{fig:type_of_slow}
\vspace{-0.3 cm}
\end{figure*}

\subsection{Four degrees of slowing in EEG}
The channel-level slowing detector provides fine-grain information about slowing in the EEG, as it determines when and where slowing occurs in the EEG. In the following, we illustrate how the channel-level detector can be utilized to detect different degrees of slowing, yielding more information for the experts to assess the EEG slowing in a patient.

Four degrees of slowing can be distinguished from the EEG slowing duration (intermittent or continuous) and localization (focal or generalized) \cite{kane2017revised, tatum2014handbook}. Following the literature, we set 20\% as the lower limit for abnormal slowing. Any channels that exhibit slowing longer than 20\% of the recording are marked as abnormal. If the number of abnormal channels is more than 50\% of the total number of channels, the EEG exhibits generalized slowing, otherwise the slowing is considered focal. Next, we compute the average percentage of slowing duration in those abnormal channels. If the percentage is over 90\%, it is continuous slowing, otherwise it is intermittent if the slowing is above 20\% and below 90\% of the recording. Usually, EEG slowing can be considered generalized if it occurs at more than half of the electrodes. However, in some special cases, it might be viewed as focal even if most electrodes exhibit slowing, e.g., a right-hemispheric slowing from an earlier surgery, and left temporal intermittent slow waves will be considered two separate focal pathologies.

We illustrate the four degrees of EEG slowing in Figures \ref{fig:degree_of_slowing} and \ref{fig:type_of_slow}. The scatterplot is divided into four quadrants at the 50\% mark on both axes to reveal four regions: continuous generalized slowing (CGS), intermittent generalized slowing (IGS), continuous focal slowing (CFS), and intermittent focal slowing (IFS). From the DLDS scatterplot, we select an EEG example for each degree of slowing and a slow-free EEG (case a to e), and plot scalp heatmaps of the percentage of slowing for each case in Figure \ref{fig:type_of_slow} to differentiate the different degrees of slowing visually from the contours.

\section{Discussion}
\label{discuss}

\subsection{Comparison of the proposed classification systems}
Excluding the LTMGH dataset, the DLDS exhibits the best overall classification performance for the three classification problems for both LOIO and LOSO CV. However, when we include the LTMGH dataset, the TDS yields better EEG-level LOSO CV results. Examining the results from the SLDS and DLDS also show that a channel-level slowing detector based on deep learning is superior to one based on shallow learning. Therefore, it is recommended to deploy the DLDS instead of the SLDS.

The results suggest that we should always recalibrate the systems to bestow superior classification accuracy. Otherwise, we train the proposed systems with EEG data from other institutes. Assuming we have EEGs from a new dataset, we should always deploy the DLDS for most cases as it is superior to the TDS. Even if channel-level annotations are unavailable for this dataset, we can deploy a slowing detector trained on other datasets and only retrain the EEG classifier on the EEGs from the target center. However, we should deploy the TDS if the EEGs are recorded under non-standard conditions that may result in distortions in the EEG. The DLDS is not suitable in this situation, as the EEGs are not generalizable. 

Therefore, we should always recalibrate the systems if possible. Otherwise, we train the proposed systems with EEGs from other institutes. The DLDS is suitable for most situations. However, when the EEGs appears to be not generalizable, we should deploy the TDS to recalibrate the system.

\subsection{Comparison of the performance of the system on various EEG datasets}
The performance of the three EEG classification systems varies across the different datasets. The classification results for the TUH dataset are consistently excellent, probably because the dataset is prepared explicitly for slowing EEG related research. Therefore, the clinical reports might be more reliable and accurate. On the other hand, the NNI and Fortis datasets were created without such specifications nor selection biases. Hence, their clinical reports may contain less reliable information regarding slowing, leading to poorer results. However, the NNI and Fortis datasets may be more in line with routine EEGs recorded and interpreted in clinical practice. The three systems perform the worst on the LTMGH dataset, although the data collection method is the same as the NNI and Fortis datasets. This is because the EEGs from LTMGH are more prone to sweat artifacts, and the skewed spectra may hurt the classification performance, as suggested by the results.

There are also various degrees of slowing across the datasets. From the TUH dataset clinical reports, most EEG slowing appears to be generalized and/or continuous. In contrast, the slowing EEGs in the NNI and Fortis dataset is often focal and/or intermittent. Generalized and continuous slowing implies a more severe neurological condition, which might be easier to detect. Therefore, the TUH dataset is expected to have more reliable classification results than the NNI and Fortis datasets. The LTMGH dataset did not specify the severity of slowing in the clinical report. However, the slowing detected for LTMGH EEGs is similar to the NNI and Fortis EEGs. Hence, the slowing EEGs from the LTMGH dataset are mostly focal and intermittent and thus should perform poorer than the TUH dataset.

\subsection{Comparison to the literature}
As far as we know, automated methods to detect pathological slowing have not yet been proposed in the literature. Instead, existing studies concentrate on detecting neurological disorders that induce slowing. In this context, most papers concern ischaemic stroke (IS), as we will briefly review in the following \cite{finnigan2016defining, sheorajpanday2011quantitative, bentes2018quantitative}.

Finnigan et al. investigated the DAR to predict the presence of IS. They proposed a threshold of $\text{DAR}=3.7$, which results in specificity and sensitivity of 100\% for detecting IS, corresponding to a 100\% classification accuracy \cite{finnigan2016defining}. However, they assessed the method on only 46 subjects (28 healthy and 18 with IS), and the data originated from only one center. 

Similarly, Sheorajpanday et al. deployed the PRI (named Delta-Theta-Alpha-Beta Ratio (DTABR) their study) to determine the presence and absence of an IS in lacunar circulation stroke (LACS) and posterior circulation stroke (POCS) \cite{sheorajpanday2011quantitative}. They reported that the PRI is not significantly different for POCS. On the other hand, they stated that $\text{PRI}<1$ was 100\% specific for the absence of recent IS, while $\text{PRI}>3.5$ was 100\% sensitive for the presence of an IS in LACS. The optimal accuracy is achieved at $\text{PRI}=1.75$, where the classification sensitivity, specificity, accuracy, and AUC are 73.0\%, 67.0\%, 71.0\%, and 0.78, respectively. However, for predicting the unfavorable outcome of IS, at $\text{PRI}=2.4$, they achieved a sensitivity, specificity, accuracy, and AUC of 100\%, 77.0\%, 83.0\%, and 0.88, respectively. They evaluated their technique on a small dataset of 60 subjects (36 subjects with POCS and 24 subjects with LACS), recorded at the same center. 

Similarly, Bentes et al. deployed two quantitative EEG indices to predict whether the post-stroke functional outcome is favorable at discharge and after 12 months \cite{bentes2018quantitative}. The alpha RP achieved a CV AUC of 0.814 and 0.852 at discharge and after 12 months, respectively. The PRI (named DTABR in the study) reached a CV AUC of 0.827 and 0.859 at discharge and after 12 months, respectively. They evaluated their methods on EEGs from 151 patients with consecutive anterior circulation ischemic stroke (112 male and 39 female), recorded from the same center.

We cannot directly compare our results with those reported in the three studies {\cite{finnigan2016defining, sheorajpanday2011quantitative, bentes2018quantitative}}, as the classification problems are different. Nonetheless, we can estimate how well our proposed system may fare in comparison. We compare the EEG-level LOSO CV results to the literature as the studies investigate EEGs from a single center. Excluding the LTMGH dataset, the DLDS yields an EEG-level LOSO CV mean sensitivity, specificity, accuracy, and BAC of 72.3\%, 91.3\%, 83.9\%, and 81.8\%, respectively, with a mean AUC of 0.851 for detecting pathological slowing. The proposed DLDS achieved better results than reported in Sheorajpanday et al. However, they are inferior to the results of Finnigan et al., while the study of Bentes et al. reported an AUC value on par with our study. Hence, our proposed DLDS achieved comparable results as compared to the literature.

However, we evaluated our proposed systems on multiple independent datasets, accounting for 613 subjects (442 and 171 subjects with slowing and no slowing, respectively) from three countries and three institutes (if we omit the LTMGH dataset). Including the LTMGH dataset, we have 1713 subjects (1143 and 570 subjects with slowing and no slowing, respectively) from three countries and four institutes. Furthermore, we assessed our systems for both LOIO and LOSO CV scenarios, while all the studies only tested their method via LOSO CV. More importantly, our systems are not restricted to stroke prediction, as it detects pathological slowing in general. Consequently, they can be applied to identify disorders that induce pathological slowing.


\subsection{Computational complexity}
We assess the processing time required for classifying a 30-minute routine EEG by the three proposed systems. The experiments were conducted in Python v3.7 with an Intel (R) Core$^{\text{(TM)}}$ i5-6500 CPU @ 3.20G Hz and a Nvidia GeForce GTX 1080 Graphical Processing Unit (GPU) with Ubuntu 16.04 as the operating system. The evaluation was executed over 100 trials, and the statistics are summarized in Table \ref{tab:computation_time}. 

The DLDS is the most computationally efficient, with a mean processing duration of around 4s, as the system runs the CNN on the GPU. The TDS and SLDS took almost 12 times longer to process the EEG. Most of the time is spent on extracting spectral features for the channel-level slowing detector. Consequently, the DLDS is fast enough for clinical applications and be operated in real-time such as monitoring in the ICU or fast triage in the emergency department, whereas the TDS and SLDS would require more efficient implementations for such purposes. 

For instance, we can parallelize the feature extraction method and extract at each channel in parallel; such a parallel scheme would lead to a drastic speed-up. Moreover, we can reduce the overlap percentage to reduce the number of segments as computation time for the channel-level features extraction module is proportional to the number of segments. However, we will have to determine the optimal overlap percentage to prevent compromising the classification performance.

\begin{tablehere}
\tbl{Processing time required for a 30-minute EEG (128Hz).\label{tab:computation_time}}
{
\scalebox{0.95}{
\centering
\begin{threeparttable}
\begin{tabular}{|c|c|c|c|} 
\hline
\textbf{System} & \textbf{TDS} & \textbf{SLDS} & \textbf{DLDS} \\ 
\hline
\begin{tabular}[c]{@{}c@{}}\textbf{Preprocessing and}\\\textbf{artifact rejection}\end{tabular} & 2.8$\pm$0.12 & 2.8$\pm$0.12 & 2.8$\pm$0.12 \\ 
\hline
\begin{tabular}[c]{@{}c@{}}\textbf{Channel features}\\\textbf{extraction}\end{tabular} & 45.3$\pm$0.54 & 45.3$\pm$0.54 & - \\ 
\hline
\begin{tabular}[c]{@{}c@{}}\textbf{Channel-level}\\\textbf{classification}\\\textbf{~(CPU + GPU)}\end{tabular} & - & 0.042$\pm$0.13 & 1.1$\pm$0.28 \\ 
\hline
\begin{tabular}[c]{@{}c@{}}\textbf{Histogram features}\\\textbf{extraction}\end{tabular} & 0.19$\pm$0.089 & ~0.007$\pm$0.002 & 0.009$\pm$0.01 \\ 
\hline
\begin{tabular}[c]{@{}c@{}}\textbf{EEG-level}\\\textbf{classification}\end{tabular} & 0.09$\pm$0.04 & 0.27$\pm$0.047 & 0.16$\pm$0.008 \\ 
\hline
\textbf{Mean total time} & 48.4 & 48.4 & 4.1 \\
\hline
\end{tabular}

\begin{tablenotes}
\setlength\labelsep{0pt}
\footnotesize
\item Time is reported as mean $\pm$ std seconds.
\end{tablenotes}

\end{threeparttable}
}}
\vspace{-0.3 cm}
\end{tablehere}

\section{Conclusions and Future work}
\label{conclude}
We proposed three automated systems to detect pathological slowing in EEG. Slowing can be detected on the channel-, segment-, and EEG-level. We evaluated the proposed systems on datasets from TUH, NNI, Fortis, NUH (only channel- and segment-level), and LTMGH (only EEG-level). The DLDS yielded the best overall classification results (excluding LTMGH): LOIO CV mean BAC of 71.9\%, 75.5\%, and 82.0\%, for the channel-, segment- and EEG-level classification, respectively, and LOSO CV mean BAC of 73.6\%, 77.2\%, and 81.8\%, for the channel-, segment-, and EEG-level classification, respectively. The TDS and SLDS approach the EEG-level performance of the DLDS with a BAC of 82\% for LOSO CV, but underperform in other situations. 

The channel and segment-level performance of the DLDS has an LOIO CV mean BAC of 71.9\% and 77.2\%, which is similar to the channel and segment IRA of the expert, which stands at 72.4\% and 82\%, respectively. This suggests that the DLDS system can detect EEG slowing channel- and segment-wise on par with a human expert. Similarly, the EEG-level results for the three systems are comparable to human experts as it lies within the range of the inter-rater agreement of 80\% for detecting IED patterns in EEG \cite{noe2020most}. At present, there are no similar inter-rater agreement studies for EEG slowing, which would be a more relevant benchmark for automated detection of EEG slowing.

To gain more insights into the automated detection of EEG slowing, we developed and assessed the histogram-based features of the EEGs. By defining the percentage of slowing and normal background, we deployed a threshold-based EEG-level classification method that yields decent accuracy. Moreover, we also define the four degrees of slowing, to illustrate that our systems can provide detailed information about slowing in the EEG, which can assist the EEG reviewing process. We distinguish four degrees of EEG slowing depending on the duration and spatial extent of the slowing, following definitions from the clinical literature. This allows us to visualize the various type of EEG slowing on the scalp, which may provide helpful information for diagnostic purposes in the future. The DLDS can evaluate a 30-minute EEG in around 4s, allowing real-time clinical applications such as continuous ICU monitoring, brain surgery, and anesthesia monitoring \cite{swisher2015baseline, hutt2018suppression}.

This study has several limitations. The proposed systems have only been tested on EEGs recorded while the subjects are awake, which may be less common in some routine EEG settings. In future work, we will train models to handle sleep EEGs. Additionally, while we manage to detect slowing in EEG, and described four possible degrees of slowing based on the duration and spatial extent, this distinction of four degrees was not validated. As the clinical reports do not contain information about the duration and spatial extent of slowing, we were not able to verify how reliably our systems can detect different types of slowing. Moreover, we did not explore how to identify the neurological disorder potentially causing EEG slowing, such as stroke, dementia, or brain injury; this could be an interesting topic for future work.

Considering that the SLDS and DLDS are trained with the annotations marked by the expert, in future work, we want to increase the number of segments for the expert to annotate in order to increase the size of the dataset for training and testing the channel-level slowing detector. Additionally, we want to collect annotations from more than one expert to develop the channel-level slowing detector based on the opinions of multiple experts. This can be achieved by implementing hard/soft voting via ensemble learning, to reduce bias and improve the reliability of the system. With more annotations from multiple experts, we will investigate the slowing inter-rater agreement between multiple experts. Lastly, we will look into newer and more powerful supervised classification algorithms such as finite element machine and dynamic ensemble algorithm \cite{pereira2020fema, alam2020dynamic}.

\section*{Conflicts of Interest} 
\noindent The authors have no disclosures to report.

\section*{Acknowledgments} 
\noindent The NUH and NNI datasets were collected under the supervision of Dr. Rahul Rathakrishnan and Dr. Yee-Leng Tan, respectively, supported by the National Health Innovation Centre (NHIC) grant (NHIC-I2D-1608138).

\section*{Appendix}

\subsection*{A. EEG segment preparation}
We prepared 1000 5s EEG segments to be annotated by an expert on both the channel- and segment-level. We included waveforms that are likely to be slow-free, in addition to waveforms that most likely exhibit slowing. The PRI is selected as a slowing measure due to its excellent segregation performance for stroke-related conditions \cite{sheorajpanday2011quantitative}. Indeed, it has been reported that the average PRI value across 19 channels yields 100\% specificity for $\text{PRI}<1$, and 100\% sensitivity for $\text{PRI}>3.5$.

Similarly, we select segments with $\text{PRI}\geq3.5$ to include highly probable slowing segments. We also include a small number of segments with PRI in-between $1<\text{PRI}<3.5$ to include segments that should contain more slow-free channels. Segments with PRI value $\text{PRI}<1$ are not selected as most channels are expected to be slow-free. The segment length is 5s to allow the segment to contain up to five periods of slowing waveform in any channel. We extracted waveforms from the TUH, NNI, Fortis, and NUH dataset to obtain annotated segments and channels and select the waveforms for annotation according to the following procedure:

\begin{enumerate}[leftmargin=*,align=left]
\item We apply the EEG preprocessing methods described in Section II and split the EEG into 5s segments and extract the average PRI values across the 19-channels.
\item We remove segments that contain noticeable artifacts or other obvious abnormalities such as seizures by visual inspection.
\item We randomly select 950 unique segments, with approximately equal numbers from the TUH, NNI, Fortis, and NUH datasets. We select a maximum of 20 segments from each EEG to include waveforms from a large variety of EEGs. Additionally, we select the segments such that 90\% of the segments have $\text{PRI}>3.5$ (highly probable slowing segment), and 10\% of the segments have $1<\text{PRI}<3.5$ (highly probable slow-free segment). Each of the highly probable slowing and slow free segments are sampled uniformly and randomly among all the extracted segments. We used this ratio as not all channels from slowing segments will contain slowing, while the majority, if not all of the channels in slow-free segments will be free of slowing.
\item Then, we select 50 random waveforms, with almost equal distribution from each dataset, and create one duplicate for each of them.
\item Finally, we combined the 950 unique segments and 50 duplicates and randomize the order of all the segments.
\end{enumerate}

One expert annotated the segments and channels in NeuroBrowser (NB) \cite{jing2016rapid}. The expert also pointed out other types of waveforms, including artifacts, ictal activities, spikes, eye blinks, K-complexes, photic stimulations, or NIL (no comments).

\subsection*{B. PRI values for events waveform}
In Table \ref{tab:unique_segment_summary_all}, we show PRI values of the events mentioned by the expert in the comments. The segments containing spikes, eye blinks, and K-complex waveforms have a much higher PRI value on average. By contrast, segments containing artifacts, ictal activity, and photic stimulation waveforms have a much lower PRI value. As a high PRI value is a feature of slowing, this implies that sharp spike waveforms can generate slow-like features, while waveforms with artifacts can appear slow-free. 

However, this can result from the high-pass filter applied in the study, where such filters can generate slow oscillations on transients or spike inputs, resulting in a high PRI value.  Nonetheless, the application of such a filter does not appear to affect the classification results significantly.

\begin{tablehere}
\vspace{-0.4cm}
\tbl{Summary of the PRI values extracted from events. \label{tab:unique_segment_summary_all}}
{
\centering
\scalebox{0.9}{
\begin{tabular}{|c|c|c|c|c|c|c|} 
\hline
\multirow{2}{*}{ \textbf{Event} } & \multicolumn{3}{c|}{\begin{tabular}[c]{@{}c@{}}\textbf{Slow-free}\\\textbf{Segment PRI} \end{tabular}} & \multicolumn{3}{c|}{\begin{tabular}[c]{@{}c@{}}\textbf{Slowing}\\\textbf{Segment PRI} \end{tabular}} \\ 
\cline{2-7}
 & \textbf{No}  & \textbf{Mean}  & \textbf{std}  & \textbf{No}  & \textbf{Mean}  & \textbf{std}  \\ 
\hline
\textbf{Artifact}  & 72 & 6.125 & 3.678 & 8 & 9.751 & 4.196 \\ 
\hline
\textbf{Ictal}  & 12 & 5.323 & 2.504 & 10 & 7.868 & 3.767 \\ 
\hline
\textbf{Spike}  & 7 & 9.691 & 6.235 & 3 & 14.714 & 10.311 \\ 
\hline
\textbf{Eye blink}  & 9 & 12 & 9.791 & 2 & 27.091 & 25.297 \\ 
\hline
\textbf{K-complex}  & 5 & 12.865 & 15.53 & 1 & 34.195 & - \\ 
\hline
\textbf{Photic}  & 0 & - & - & 1 & 3.897 & - \\ 
\hline
\textbf{NIL}  & 460 & 6.302 & 4.072 & 311 & 17.335 & 19.828 \\ 
\hline\hline
 \textbf{All}  & 565 & 6.449 & 4.455 & 335 & 16.926 & 19.341 \\
\hline
\end{tabular}
}}
\vspace{-0.4 cm}
\end{tablehere}

\bibliographystyle{ws-ijns}

\end{multicols}
\end{document}